\documentstyle[epsfig,referee]{mn} 
\date{received,  accepted}  
\title[The SCUBA 8-mJy Survey]{The SCUBA 8-mJy survey - II:  
Multiwavelength analysis of bright sub-mm sources} 
 
\author[Fox et al] 
{M.\,J.\, Fox $^{\! 1}$, A. Efstathiou $^{\! 1}$, M.\, Rowan-Robinson $^{\! 1}$, J.S.\, Dunlop $^{\! 2}$, S. Scott $^{\! 2}$, \and S. Serjeant $^{\! 1, 7}$, R.G. Mann $^{\! 1}$$^{\! ,2}$,  S.\, Oliver $^{\! 1, 8}$, R.\,J.\, Ivison $^{\! 3}$, A. Blain $^{\! 4, 9}$, O. Almaini $^{\! 3}$, \and D. Hughes $^{\! 5}$, C.J. Willott $^{\! 6}$, M. Longair $^{\! 4}$ ,  A. Lawrence $^{\! 2}$, J.\,A.\,Peacock $^{\! 2}$ 
\vspace*{3mm}\\ 
$^{1}$ Astrophysics Group, Blackett Laboratory, Imperial College of Science Technology \& Medicine, Prince Consort Rd.,London.SW7 2BW\\ 
$^{2}$ Institute for Astronomy, University of Edinburgh, Royal Observatory, Blackford Hill, Edinburgh, EH9 3HJ\\ 
$^{3}$ UK ATC,  Royal Observatory, Blackford Hill, Edinburgh, EH9 3HJ\\ 
$^{4}$ Institute of Astronomy, Madingley Road, Cambridge, CB3 0HA\\ 
$^{5}$ Instituto Nacional de Astrofisica, Optica y Electronic (INAOE), Apartado Postal 51 y 216, 72000 Puelba, Pue., Mexico\\ 
$^{6}$ Astrophysics, Department of Physics, Keble Rd, Oxford OX1 3RH\\ 
$^{7}$ Unit for Space Sciences and Astrophysics, School of Physical Sciences, University of Kent, Canterbury, Kent, CT2 7NZ, UK\\
$^{8}$ School of Chemistry, Physics \& Enviromental Science, University of Sussex, Falmer, Brighton BN1 9QJ, UK\\
$^{9}$ Astronomy Department, California Institute of Technology, Pasadena, CA91125, USA}  
\date{Accepted ...  Received ... } 
 
\pagerange{000--000} 
 
\begin{document} 
 
\maketitle 
 
\begin{abstract} 
We present the results of a multi-wavelength study of the 19 most significant  
sub-mm sources detected in the SCUBA 8-mJy survey.  
As described in Scott et  
al. (2001), this survey covers $\simeq 260$ arcmin$^2$ using the  
sub-millimetre camera SCUBA, to a limiting source detection limit 
$S_{850\mu m} \simeq 8$~mJy. One advantage of this relatively bright 
flux-density limit is that accurate astrometric positions are potentially 
achievable for every source using existing radio and/or mm-wave  
interferometers. However, an 
associated advantage is that SED-based redshift constraints should be 
more powerful than in fainter sub-mm surveys. Here we therefore exploit  
the parallel SCUBA 450$\mu m$ data, in combination with existing radio  
and ISO data at longer and shorter wavelengths to set constraints on the 
redshift of each source. We also analyse new and existing optical and  
near-infrared imaging of our SCUBA survey fields to select potential  
identifications consistent with these constraints. Our  
derived SED-based redshift constraints, and the lack of statistically significant   
associations  
with even moderately bright galaxies allow us to conclude  
that all 19  
sources lie at $z > 1$, and at least half of them  
apparently lie at $z > 2$.

\end{abstract} 
 
\begin{keywords} 
cosmology: observations ; galaxies: starburst ; galaxies: distances and redshifts ; galaxies: evolution  
\end{keywords} 
 
\section{Introduction}   
Even prior to the advent of the first major sub-millimetre  
surveys it was anticipated that, if  
substantial numbers of sources were to be uncovered by surveys conducted 
at 850$\mu m$, the  
vast majority of these would most probably lie at high redshift $z > 1$ 
(Blain \& Longair 1996, Hughes \& Dunlop 1998). 
This is a simple consequence of the realization that the present-day IRAS  
luminosity function needs to be subjected to strong cosmological  
evolution (comparable to that displayed by powerful AGN out to $z \simeq 2$) 
in order to yield a significant  
number of sources in currently feasible 850$\mu m$ surveys. 
 
However, while a series of surveys with the Submillimetre Common User  
Bolometer Array (SCUBA) on the James Clerk Maxwell Telescope (JCMT) 
have now confirmed the existence 
of large numbers of sub-mm sources (Smail et al. 1997, Hughes et al. 1998,  
Barger et al. 1998, Eales et al. 1998, Blain et al. 1999 and  
Barger et al. 1999a) actually measuring or even  
constraining the redshift distribution of this important extragalactic  
population is proving  
extremely difficult. There are a number of reasons for this, perhaps the most  
obvious of which is that very few of these sources transpire to be associated 
with observable AGN emission (Fabian et al. 2000, Hornschemeier et al. 2000, Barger et al. 2001a, 2001b). In fact, to date spectroscopic  
redshifts have only been measured for 3 of the bright sub-mm sources  
uncovered by blank-field SCUBA surveys, in two cases (SMMJ02399-0136 at  
$z = 2.8$ and SMMJ02400-0134 at $z=1.1$) aided by the (apparently)  
rare occurrence of detectable AGN activity 
(Ivison et al. 1998, Soucail et al 1998). 
 
Nevertheless, the situation is not quite as hopeless as is sometimes  
portrayed. In particular, while it is clear that the ideal of  
spectroscopically-determined redshifts (via optical, maser line (Townsend et al 2001), IR, or CO  
mm-wave spectroscopy) for  
substantial numbers of SCUBA sources remains a distant  
goal, much effort has been invested in refining  
techniques of redshift estimation which can be implemented  
with current instrumentation (see Dunlop (2001) for an overview). 
 
In practice 4 key steps can be identified along the route towards establishing 
an unambiguous redshift for a sub-mm source. These are: 
 
\begin{enumerate} 
 
\item{Establish an allowed redshift range consistent with the  
observed radio-to-infrared spectral energy distribution (SED) of the source.} 
 
\item{Identify possible candidate optical/IR counterparts consistent  
both with the position of the SCUBA source and with  
the SED-based redshift constraints.} 
 
\item{Establish which (if any) of the potential optical/IR identifications  
is the correct one 
through improved astrometry provided by deep radio or mm interferometry.} 
 
\item{Given a trustworthy optical/IR identification, 
measure its spectroscopic redshift.} 
 
\end{enumerate} 
 
It is already clear that attempting to short-circuit this sequence and jump 
straight to step 4 ({\it i.e.} measure a redshift for all potential optical/IR  
counterparts) not only represents very expensive use of valuable large  
telescope time, but can produce potentially misleading   
results (Barger et al. 1999b). Indeed, given the faintness and redness of  
some of the optical/IR counterparts, there is a serious possibility 
that spectroscopic redshifts may not prove measurable for a substantial 
fraction of SCUBA  
sources until the advent of NGST and/or ALMA. 
 
It is therefore important to recognize the value of steps 1 to 3 in the above 
sequence, and to attempt to maximise the undoubted potential of such 
currently-feasible measurements for establishing the basic nature of the  
sub-mm source population. 
 
To date the effectiveness of step 1 ({\it i.e.}   
SED-based redshift constraints)  
has been hampered by the lack of sufficiently bright sub-mm sources 
revealed by existing surveys. For example, only three 
of the 850$\mu m$ sources uncovered by the SCUBA surveys of the HDF-N proper 
(Hughes et al. 1998, Serjeant et al. 2001) and of the 14hr-field of the CFRS  
(Eales et al. 2000) 
have $S_{850} > 5$mJy, and obtaining complementary detections 
of the fainter sources at either 450$\mu m$ or 1.4 GHz has, unsurprisingly,  
proved to be extremely difficult (Eales et al. 2000). 
 
Fortunately, the recently-completed `8-mJy' SCUBA survey has transformed  
this situation, yielding 36 sources with $S_{850} > 5$ mJy and S/N $> 3.5$ 
(Scott et al. 2001). 
In this paper we report the first results of attempting steps 1 and 2 for the  
19 most significant ($> 4 \sigma$) of these sources. 
An important feature of this new SCUBA-selected sample is that 
all of the sources are bright enough to be detectable  
with existing radio and/or mm interferometers (e.g. Downes et al. 1999,  
Gear et al. 2000).  
Thus, ultimately, we would  
anticipate that step 3 in the above procedure can also be completed  
for the bulk of this new sub-mm sample (Lutz et al. 2001, Ivison et al. 2001). 
 
Here we focus on what can be deduced about these sources from the existing 
multi-frequency data available for the 8-mJy survey fields. 
This survey has the advantage of deep multiwavelength data from the  
European Large Area ISO Survey (ELAIS) project (Oliver et al. 2000) for  
50$\%$ of the total area at 7, 15, 90 and 175$\mu$m.  The  
remaining 50$\%$ is covered with other ISO observations.  There also  
exists a wealth of available data in the $I$- and $R$-band  
(Willott et al. 2001), at 
X-ray wavelengths  
(Hasinger et al. 1998, Schmidt et al. 1998, Lehmann et al. 2000,  
Lehmann et al. 2001) and at 1.4~GHz (de Ruiter et al. 1999, Ciliegi et al.  
1998).  We have also now acquired deep $K$-band imaging of the central 
regions of both fields, using UFTI on UKIRT and INGRID on the WHT. 
 
The paper is organized as follows. 
In Section 2 we present and analyse the parallel  
450$\mu m$ SCUBA survey images. In Section 3 we then combine the resulting  
detections/limits with existing radio, far-infrared  
and mm-wave data to determine 
SED-based redshift constraints for the 19 most significant  
850$\mu m$ sources.  
In Section 4 we exploit deep optical and near-infrared imaging of  
our SCUBA fields to detect and quantify the probability of candidate  
identifications. Finally in Section 5 we place the main results of this  
work in context, discuss the significance of our 
principal findings, and highlight the importance of forthcoming 
deeper multi-frequency observations of the 8-mJy  
SCUBA survey. 
 
\begin{figure*} 
 \centering 
    \vspace*{10cm} 
    \leavevmode 
  \caption[]{{\bf This figure can be found at http://astro.ic.ac.uk/~mfox/paperII\_figures/lockman\_450.ps.gz} The 450$\mu$m signal to noise image of the Lockman Hole region convolved with the full beam, with the locations of the 850$\mu$m detections marked 
by circles.  The beamsize of JCMT at 450$\mu$m is 7.5$^{\prime \prime}$  
producing a higher resolution image than at 850$\mu$m.  LH850.1 and LH850.11 
 have 450$\mu$m counterparts and the remaining 850$\mu$m sources have  
450$\mu$m fluxes consistent with a non-detection. }\label{lh_450} 
\end{figure*} 
 
\begin{figure*} 
 \centering 
    \vspace*{10cm} 
    \leavevmode 
  \caption[]{{\bf This figure can be found at http://astro.ic.ac.uk/~mfox/paperII\_figures/elaisn2\_450.ps.gz} The 450$\mu$m signal to noise image of the ELAIS N2 region convolved with the full beam, with the locations of the 850$\mu$m detections marked by circles.  The  
two  
most significant sources, N2850.1 and N2850.2 have solid 450$\mu$m detections  
and the remaining 850$\mu$m sources have 450$\mu$m fluxes consistent with a  
non-detection.  The source N2450.1 is the one potentially 
real 450$\mu$m source in the map with no significant 850$\mu$m counterpart  
(see text for discussion).}\label{n2_450} 
\end{figure*} 
 
\section{Parallel 450-micron SCUBA imaging} 
 
\subsection{450$\mu m$ maps} 
 
The 8-mJy survey covers a total area of approximately 260 arcmin$^2$, 
divided roughly evenly between two fields; the Lockman Hole East, and one of  
the ELAIS survey  
regions in the northern sky, ELAIS N2. These two survey areas were  
selected for their low galactic cirrus emission and the extent of  
pre-existing multiwavelength data. As reported by Scott et al. (2001) 
both survey fields have been imaged at $\lambda = 850 \mu m$ with SCUBA to a 
typical r.m.s. noise level of $\sigma_{850} \simeq 2.2$~mJy,  
yielding 19 sources 
with S/N $>$ 4, 38 sources with S/N $>$ 3.5, and 72 sources with $S/N > 3$.  
The flux densities of the 19 most significant  
850$\mu m$ sources which are the focus of this multi-frequency analysis  
are restated here for ease of reference in Table 1. 
 
Because SCUBA observes simultaneously at  
450$\mu$m and 850$\mu$m we have also 
obtained parallel 450$\mu m$ images of these two survey fields. 
The 450$\mu$m observations are inevitably of poorer sensitivity because of  
the lower atmospheric transmission and lower aperture efficiency, and are also 
more difficult to calibrate reliably.  The atmospheric opacity at 450$\mu$m was typically less than 1.8.  Mars and Uranus were used as primary calibrators with CRL618, OH231.8 and CRL2688 as secondary calibrators and were observed using a 30'' chop throw identical to the survey strategy.  At 450$\mu$m we find a typical calibration error of 20$\%$.  
Despite this large uncertainty in the calibration the parallel 450$\mu$m data   
are of considerable value due to the fact that  
the flux-density ratio $S_{850}$:$S_{450}$ is a strong function  
of redshift. This is because the grey body spectrum  
produced by a dust-enshrouded starburst galaxy rises as steeply as  
$f_{\nu} \propto \nu^{3-4}$ in the rest-frame sub-mm, 
but gradually flattens at shorter wavelengths, turning over  
at $\lambda \simeq 100\mu m$.  
Consequently 450$\mu m$ detections of very bright and/or  
low-redshift  
850$\mu m$ sources are frequently achievable (given good atmospheric  
transparency), and even when detections are not achieved  
the resulting limits on 850/450 colour provide a valuable (albeit  
temperature sensitive) limit on source redshift.  
Furthermore, the  
shortwave array in SCUBA has 91 pixels (cf. 37 at 850$\mu$m) which,  
coupled with the smaller beamsize of the 15-m JCMT at  
450$\mu m$ (FWHM 7.3 arcsec), results in a higher resolution  
map which is thus less subject to the potential effects of source confusion. 
 
In Figures \ref{lh_450} and \ref{n2_450} we present the 450$\mu m$ images  
of the Lockman Hole and ELAIS N2 areas for which Scott et al. (2001) have  
presented maps at 850$\mu m$. We estimate that the mean 3-$\sigma$ limiting 
depths of these maps are $S_{450} \simeq 65$~mJy for the Lockman Hole area and  
$S_{450} \simeq 50$~mJy for the ELAIS N2 field.  However, we emphasize that  
these sensitivities can vary by up to 80$\%$ across the maps. 
   
\subsection{Source extraction} 
 
As discussed by Scott et al. (2001) we have reduced both the 850$\mu m$  
and 450$\mu m$ data using the IDL pipeline developed by Serjeant et al. (2001) 
in order to produce uncorrelated signal and noise images. 
This allows the use of maximum-likelihood source-extraction  
techniques as discussed by Serjeant et al. (2001) and Scott et al. (2001). 
Application of these source-extraction methods to the 450$\mu m$ images  
results in one source with S/N $> 4$ and 15 sources with S/N $> 3.5$.  
 
Four of these 450$\mu m$ sources coincide (to within the positional errors)  
with 850$\mu m$ sources extracted by Scott et al. from the longer wavelength 
images, and for this reason are almost certainly real. However, it is doubtful  
that any of the other purely 450$\mu $m-selected `sources' (those without counterparts at 850$\mu$m) can be believed. 
The reason for this is that  
while the smaller beamsize at 450 $\mu m$ means that confusion is  
a less serious problem than at 850 $\mu m$, the much larger number of beams  
in the 450 $\mu m$ maps ($\sim 5000$ across the two maps)  
means that $\sim $22 false `sources' with apparent S/N $> 3.0$  
are expected in these images purely on the basis of random noise. 
In fact simulations of the 450$\mu m$ images of the sort undertaken by  
Scott et al. (2001) at 850$\mu m$ indicate that, at most, only the one  
ELAISN2 source  
with S/N $>$ 4 can be seriously considered as a possible new 450$\mu  
m$-selected source.  At this level of significance the expected number of false sources drop to $<1$. 
The position of this possible source, N2450.1, is marked in Figure 2. 
  
\subsection{450$\mu m$ measurements of 850$\mu m$ sources} 
 
Given the relative sensitivities of the two SCUBA arrays under moderately  
good observing conditions, the failure of the 450$\mu m$ image to reveal 
any compelling new sub-mm sources is not really surprising. The real value  
of these data is therefore for quantifying the 450$\mu m$  
flux density of the known reliable 850$\mu m$ sources. 
 
The positions of 19 significant 850$\mu m$ sources reported by  
Scott et al. are overlaid on the 450$\mu m$ images shown in Figures 1 \& 2. 
In fact, 4 of the 850$\mu m$ sources are detected in the 450$\mu m$ maps 
with S/N $>$ 3, as judged by the most significant 450$\mu m$  
peak found within 6 arcsec of each nominal 850 $\mu m$ position. 
These detections  
should be taken seriously since although 
(as discussed above) several spurious 3$\sigma$ `sources' are expected  
in these maps due to random statistics, the probability of a  
spurious $>3 \sigma$ 450$\mu m$ detection occurring within 6 arcsec of a known 
850$\mu m$ source is very low. 
 
These 450$\mu m$ detections are listed in column 9 of Table 1. For the  
remaining 13 850$\mu m$ sources indicated in Figures 1 and 2 we give  
conservative 3-$\sigma$ upper limits on $S_{450}$ in Table 1. 
Both the detections and these upper limits are utilised to derive  
SED-based redshift estimates/constraints in the next section.

 \begin{table*} 
\footnotesize 
\centering 
\caption[]{Table of flux densities and magnitudes (using a 
 1.5$^{\prime \prime}$ radius aperture) of SCUBA detections and  
possible counterparts.  The numbers in parentheses refer to the  
individual objects indicated in Tables 3 and 4. \label{flux_table}}    
\begin{tabular}{lccccccccc}\hline  
Catalogue & $S_{X}$$^{\dagger}$ 	&$m_{I}$ &$m_{R}$  	&$m_{K}$ & $S_{7\mu m}$& $S_{15\mu m}$&  $S_{850}$ & $S_{450}/$mJy & $S_{1.4~GHz}/$mJy  \\ 
    Name &			&	&		&	&/mJy		&/mJy				&	/mJy	& \footnotesize{(or 3-$\sigma$ limit)}	& \footnotesize{(or 4-$\sigma$ limit)}	 \\[0.25ex]\hline 
 
LH850.1 $^a$ & $<$3   	&  $>$27.4$^*$	 &		&20.78$\pm$0.03	&$<$0.1	&$<$0.1& 10.5$\pm$1.6  & 25$\pm$7  &  0.062$\pm$0.013$^{**}$   		\\ 
 
LH850.2 &$<$12  	&    22.9$\pm$ 0.1	&	&			&&&10.9$\pm$2.4 & $<$40     &  $<$0.28  	  \\  
  
LH850.3 &$<$12 		&23.4$\pm$0.1	&		&			&&& 7.7$\pm$1.7  &  $<$22   & $<$0.16     		\\ 
    
LH850.4 &$<$12 		&(1) 22.47$\pm$ 0.1 &		&  21.02$\pm$0.20	&&& 8.3$\pm$1.8 &  $<$33   & $<$0.12   			 	\\  
      &	$<$12 		&(2) 22.59$\pm$0.1  &		&  19.27$\pm$0.07	&&&    ``         &  ``   &   $<$0.12			 	\\  
	&$<$12 		&(3) $>$24.5   &		&  20.86$\pm$0.18	&&&    ``         &  ``   &   $<$0.12			 	\\  
LH850.5 &$<$12           & $>$24.5     &		&			&&& 8.6$\pm$2.0 &  $<$26  &   $<$0.16         \\  
 
LH850.6 &$<$12		&23.04$\pm$0.10	&		&			&&& 11.0$\pm$2.6 &  $<$40   & $<$0.12    				\\ 
 
LH850.7  &$<$12      &    23.5$\pm$0.1  &		&			&&& 8.1$\pm$1.9 & $<$46   & $<$0.24  			    \\ 
 
LH850.8 $^b$   &36 		&  (1) 20.72  &21.8 &  17.98$\pm$0.01		&&& 5.1$\pm$1.3  & $<$21     &   $<$0.12 \\ 
         &  $<$12   	&  (2) 21.78  &22.4  	&19.64$\pm$0.01		 	&&& `` 	  & ``     &  0.13$\pm$0.03   \\   
	 &  $<$12   	&   (3) $>$24.5   &  	&20.22$\pm$0.02	 		&&& `` 	  & ``     &     $<$0.12		\\ 
 
LH850.11 &$<$12  	&23.5$\pm$0.1	&		&			&&&13.5$\pm$3.5 & 77$\pm$20 &$<$0.16   			   \\ 
 
LH850.12  &$<$12	&  (1) 22.71$\pm$ 0.07	&	&			&&&  6.2$\pm$1.6   &  $<$27   &0.29$\pm$0.04$^c$  \\       
        &$<$12		&  (2) 23.3$\pm$ 0.13	&	&			&&&  ``	   &  ``   & $<$0.16 \\ 
 
LH850.14 &$<$12  		&      $>$24.5	&	&			&&&9.5$\pm$2.8 & $<$70     & $<$0.24    		     \\ 
     
LH850.16 &$<$12		& 22.68$\pm$0.07	&	&			&&& 6.1$\pm$1.8  &  $<$27    & $<$0.12    			\\  
  
LH850.18  &$<$12 	&(1) 23.08$\pm$0.11	&	&			&&& 4.5$\pm$1.3 &  $<$16   &$<$0.12    					      \\ 
      &			&(2) 23.35$\pm$0.14	&	&			&&&'' 	 &  ``	   & ``    \\[0.5ex] \hline

N2850.1  &		&(1) 22.7$\pm$0.02 & 23.40  $\pm$  0.01	& &$<$1&$<$2	& 11.2$\pm$1.6 &   23$\pm$7   &$<$0.30     	    \\ 
 	&		&(2)$>$26 	&  26.46 $\pm$ 0.20 	&&$<$1&$<$2	& ``	 &   ``   &   ''    	    \\ 
     	&		&(3)$>$26 	& 25.95 $\pm$ 0.12	&&$<$1&$<$2	& `` &   ``   &''   	    \\   
 
N2850.2 $^d$ &		&(1) 24.76$\pm$0.10	&	&20.57$\pm$0.04&$<$1&$<$2& 10.7$\pm$2.0 & 35$\pm$10  &$<$0.30 \\	    	 
       &		&(2) 24.82$\pm$0.10	&		&20.64$\pm$0.03& $<$1&$<$2& `` & ``  &''\\ 
 	&		&(3) $>$26 	&		&20.68$\pm$0.03&$<$1&$<$2& ``	 & ``   &'' \\ 
	&		&(4) 25.56$\pm$0.21	&		&20.96$\pm$0.03           &$<$1&$<$2& ``  & ``   &''\\ 
	&		&(5) 24.66$\pm$0.09	&		&$>$21.5           &$<$1&$<$2& ``  & ``   &''\\

N2850.3  &		&(1)25.15$\pm$0.15	& 26.54$\pm$0.19	& $>$21.5   &$<$1&$<$2& 8.5$\pm$2.2 &  $<$19   &	$<$0.30 	   			\\  
       &		&(2)$>$26 		& 25.93$\pm$0.12	&$>$21.5  &$<$1&$<$2& `` &  ''   &	``	   			\\  
  	&		&(3)$>$26 		&   $>$27.0 		&  21.06$\pm$ 0.04     &$<$1&$<$2& `` &  ''   &	``	   			\\

N2850.4 $^e$&		&(1)$>$26 	& 26.26 $\pm$0.17	&$>$21.5	&$<$1&$<$2& 8.2$\pm$1.7 & $<$34  &	$<$0.33   		\\ 
       &		&(2)22.5$\pm$0.01	& 22.68$\pm$0.10	& 18.58$\pm$0.009&$<$1&$<$2& ``  & ``  &	`` 		\\ 
  	&		&(3)$>$26	& 25.01$\pm$0.13	& $>$21.5		&$<$1&$<$2& `` & `` &''   		\\ 
N2850.5  &		&25.04$\pm$0.13	&25.33 $\pm$0.07	&&$<$1&$<$2& 8.5$\pm$2.2 & $<$18  &  $<$0.25      			\\

N2850.7  &		&(1)23.51$\pm$0.04	&24.10 $\pm$0.03	&19.93$\pm$0.03 &$<$1&$<$2& 9.0$\pm$2.4  & $<$32   & $<$0.25	    			\\ 
       &		&(2)$>$26	&25.63$\pm$0.97	& $>$21.5  &$<$1&$<$2& ``  & ``   & ``	    			\\ 
   	&		&(3)24.72$\pm$0.10	&    $>$27.0 			&20.30$\pm$0.03 &$<$1&$<$2& ``  & ``   & ``	    			\\ \hline 
\end{tabular} 
\\(a) Additional photometry: $S_{1.2mm}=3.8\pm0.5$~mJy,  
$S_{1.26mm}=3.03\pm0.56$~mJy, $S_{3.3mm}<0.6$~mJy (Lutz et al. 2001)\\  
(b) IRAM 30-m detection at $S_{1.2mm}=1.56\pm0.32$~mJy.  Source blended with LH850.1. Catalogued as $<$9 arcsec in size at 1.4~GHz (de Ruiter et al. 1997)\\ 
(c) Unresolved at 1.4~GHz. (de Ruiter et al. 1997)\\ 
(d) R-band image heavily contaminated by diffraction spike of nearby star.\\ 
(e) IRAM 30-m detection $S_{1.2mm}=2.59\pm0.42$~mJy.  \\ 
$\dagger$ 0.5-2.0keV /$10^{-16}{\rm erg s^{-1} cm{^-2}}$\\ 
$^*$ 1$^{\prime \prime}$ diameter aperture (Lutz et al. 2001)\\ 
$^{**}$ Ivison et al. 2001 \\ 
\end{table*} 
 
\section{SED-based redshift constraints} 
 
\subsection{450/850$\mu$m flux-density ratio}\label{short} 
 
\begin{figure*} 
 \centering 
    \vspace*{13cm} 
    \leavevmode 
    \includegraphics{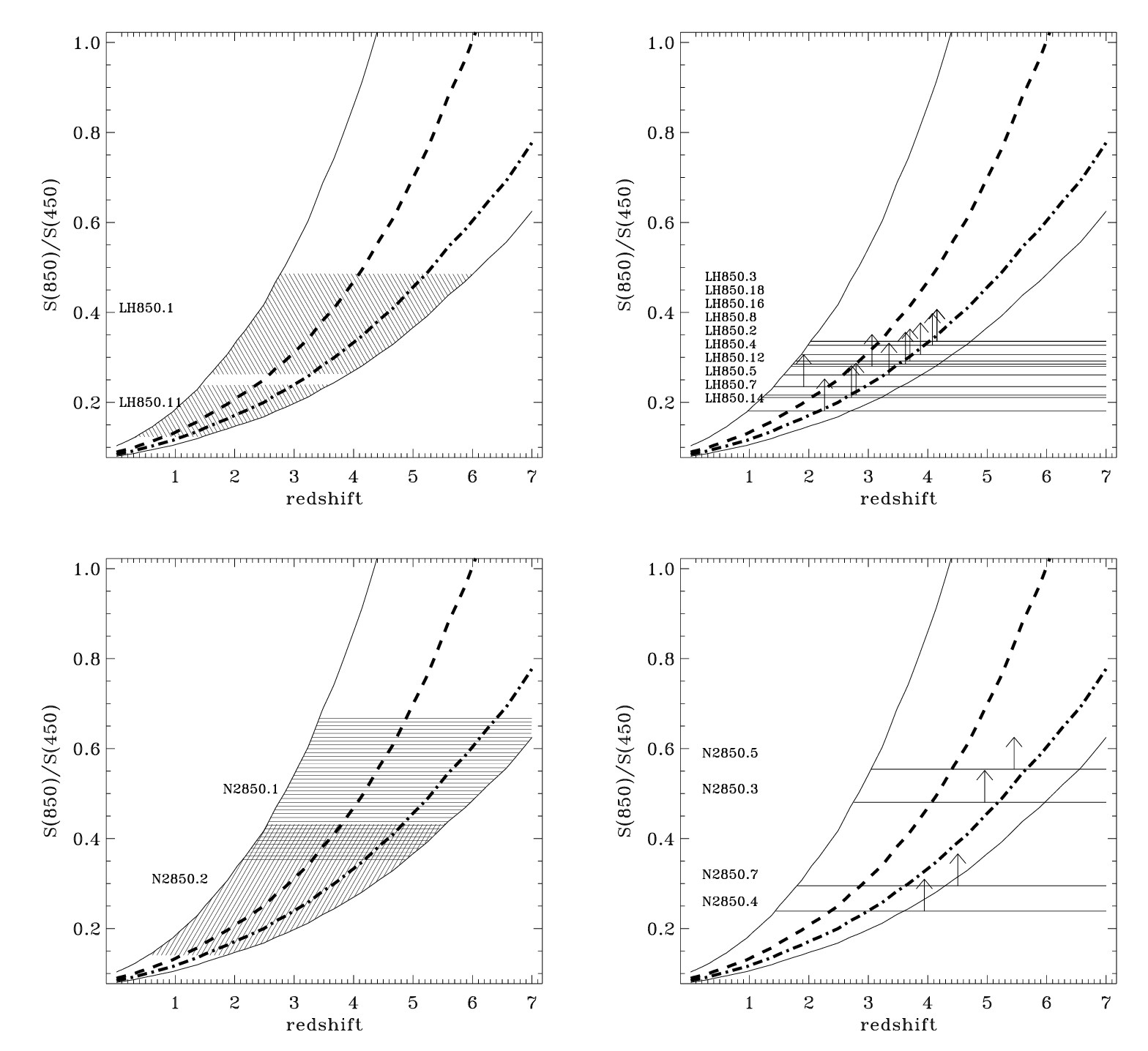} 
  \caption[]{$S_{850}/S_{450}$ colour-redshift constraints for  
the 850$\mu m$ sources in the  Lockman Hole E (top panels) and  
ELAIS N2 (bottom panels) survey fields.  The figures on the left illustrate the 
constraints derived for the four 850$\mu$m-selected sources which have  
also been detected at 450$\mu$m. The locus bounded by the two  
solid curves indicates how $S_{850}/S_{450}$ is predicted  
to vary with increasing $z$ for the range of starburst models  
described in the text, while the dashed and dot-dashed lines are  
the $S_{850}/S_{450}$ colour-redshift relations for Arp220 and M82. 
The shaded regions thus indicate 
the range of possible redshifts for these four sources, consistent  
with the 1-$\sigma$ errors on their observed 
sub-mm flux-density ratios. The figures on the right then illustrate  
what redshift limits can be derived for the remaining  
15 850$\mu m$ sources which have only 450$\mu$m upper limits.  
The ranges and limits on source redshifts derived from the comparison 
illustrated here are tabulated in column 4 of Table 2. \label{z_850_450}} 
\end{figure*} 
 
\begin{table*} 
\footnotesize 
\centering 
\caption[]{Current redshift information for the 19 most  
significant 850$\mu m$ sources from the 8-mJy SCUBA survey.  
Column 1 gives source names as defined in Scott et al. (2001). 
Column 2 lists spectroscopic redshifts for possible counterparts,  
currently available for two sources both for LH850.8  
(Lehmann et al. 2001), see text for discussion of this object.  Column 3 gives estimated  
redshift ranges and limits based on the redshift dependence  
of radio-submm  
spectral index ($\alpha_{20cm}^{850\mu m}$; Carilli \& Yun 1999, 2000).  
The $S_{1.4GHz}$ upper limits are 4-$\sigma$ for the Lockman Hole  
(de Ruiter et al. (1997)) and 5-$\sigma$ for the  
ELAIS N2 (Ciliegi et al. 1999) sources.  
Column 4 gives the ranges and limits on redshift derived  
from sub-mm colour  $S_{850\mu m}/S_{450\mu m}$ as illustrated 
in Figure 3. Column 5 gives the crude lower redshift limits which  
follow from the failure to detect the SCUBA sources in 175$\mu m$  
ISOPHOT maps. Finally, Column 6 gives the redshift ranges for 4 sources 
allowed by their detection at 1.2mm with the Mambo array at IRAM, 
providing interesting {\it upper} limits on redshift for these objects. 
The numbers in parentheses are the source references from  
Table 3. \label{redshifts}} 
\begin{tabular}{lcllll}\hline 
Name  & $z_{spect}$ &   $z(\alpha^{850\mu m}_{20cm})$ & $z(S_{850\mu m}/S_{450\mu m})$  & $z(S_{850\mu m}/S_{175\mu m})$  & $z(S_{850\mu m}/S_{1.2mm})$   
\\[0.25ex]\hline 
   
LH850.1 	&-&  \phantom{$>$}$2-4$& \phantom{$>$}$2-6$   	&   $>$1	&   $0.5-5$		\\ 
LH850.2 	&-&$>$1.5	&  $>$1  	&   -   	&   -	  \\ 
LH850.3	&-	&$>$1.5		&  $>$2    &     -	&        - 	 \\ 
LH850.4 	&-&$>$1.5	&  $>$1.5    &   -	&   -    \\ 
LH850.5		&-&$>$1.5	&  $>$1.5    &   -	&   -    \\ 
 
LH850.6	&-	&$>$1.5		&  $>$2     &    -	&   -    \\ 
LH850.7		&-&$>$1.5	&  $>$1    &     -	&   -  \\ 
LH850.8 &-	& $>$1.5 	&  $>$1	   &    $>$1 	&  $0-3$   	\\ 
LH850.8 (1)&0.974  & -	   &-	   &      -	&   -	\\ 
LH850.8 (2)&0.685  & -	   &-	   &      -	&   -	\\ 
LH850.11 	&-&$>$1.5	&  \phantom{$>$}$0.5-3$     &   - 	&   -  \\ 
LH850.12	&-&  \phantom{$>$}$0.5-2$	&  $>$1     &    - 	&   - 		\\ 
LH850.14 	&-&$>$1.5	&  $>$1     &    -	&   -  \\	 
LH850.16	&-&$>$1.5	&  $>$2	     &   - 	&   -      \\  
LH850.18 	&-&$>$1.5	&  $>$2      &   - 	&   -  \\ \hline 
        
N2850.1 	&-&$>$1.5	& \phantom{$>$}$2-7$         &  $>$1 	&   -   \\  
N2850.2 	&-&$>$1		& $1-5$    	& $>$1 	&   - \\  
N2850.3 	&-&$>$1.5	& $>$2.5       & $>$1   	&     -   \\ 
N2850.4 	&-&$>$1.5	& $>$1.5     & $>$1   	&  $0-3$   	\\ 
N2850.5 	&-&$>$1		& $>$3	  & $>$1 	&   -	\\ 
N2850.7 	&-&$>$1.5	& $>$1	  & $>$1 	&   -	\\ \hline 
 
\end{tabular} 
\end{table*}

For the estimation of redshift based on sub-millimetre flux-density ratio  
we have considered 
a range of model spectra similar to local ultraluminous infrared galaxies  
(for a review see Sanders \& Mirabel 1996).   
Plotted in Figure \ref{z_850_450} are the   
predicted sub-mm flux-density ratios as a  
function of redshift for a range of model SEDs  
from Efstathiou et al. (2000) produced by varying  
optical depth ($\tau_{v} =50-200$)  
and starburst duration (1.7 to 72 Myrs). This ensemble of model SEDs in effect 
spans a  
wide temperature range, from `Milky Way-like' dust temperatures ($\sim$20 K)  
through those typically found in local luminous infrared galaxies ($\sim$35 K,  
Dunne et al. 2000) and extending up to the higher temperatures displayed by  
some ULIRGS and HLIRGs ($\sim$50 K, Farrah et al. 2001, Dunne et al. 2000).   
For those sources with 450$\mu$m detections the  
$S_{\rm 850}/S_{\rm 450}$ colour  
constraint provides both an upper and lower redshift limit based on comparison 
of the upper and lower 1-$\sigma$ error bounds on the measured flux-density  
ratio with the  
locus of colour as a function of $z$ predicted by the model ensemble. 
In the majority of cases the 850$\mu m$ sources remain undetected in  
the 450$\mu m$ map, in which case we simply derive a lower limit  
on 850/450 colour from the 3-$\sigma$ upper limit on $S_{450}$ given in Table  
1, and hence derive a conservative lower limit on $z$ via comparison 
with the model colour locus shown in Figure 3. 
 
The resulting inferred redshift ranges and limits for the 19 850$\mu$m  
sources are summarised in column 4 of Table 2.   
The ranges are wide, and the limits almost certainly conservative 
due to the wide range of model SEDs used in this analysis. However, 
the constraints are still sufficiently strong to conclude that 
the vast majority of the sources lie at $z>1$, while at least  
half have redshifts $z > 2$.  
 
Coadding the 450$\mu$m limits and using the mean 850$\mu$m value  
for the sample yields a mean redshift lower limit for the sample of 
$\langle z_{lim} \rangle = 1.5$, using the most conservative model  
SED (ie the upper curve in Figure \ref{z_850_450}).   
Re-computing this number with an Arp220-type SED  
(dashed line in Figure \ref{z_850_450}) yields  
$\langle z_{lim} \rangle = 2.3$.

\subsection{850/1200$\mu$m flux-density ratio} 
 
LH850.1, LH850.8 and N2850.4 have been observed at 1.2mm using the MAMBO instrument  
(Kreysa et al. 1998) at the IRAM 30m telescope. Data reduction was  
performed using the procedure 
described by Baker et al. (2001). The central photometric  
bolometer was centred on the 850$\mu$m position and yielded detections  
at 1.2mm for all three objects within the 11-arcsec IRAM beam.  These  
longer wavelength measurements are of use because they can exclude  
very high redshifts on the basis of the resulting  
$S_{\rm 850\mu m}/S_{\rm 1.2mm}$ flux ratios. Using the same wide  
range of SEDs as discussed above leads to the conclusion that 
the measured $S_{\rm 850\mu m}/S_{\rm 1.2mm}$ value for LH850.1 is consistent with  
the redshift range $0.5 < z < 5$ taking the upper and lower  
1-$\sigma$ error on the observed ratio. On the same basis,  
LH850.8 and N2850.4 are consistent with $0 < z < 3$. These albeit 
broad redshift ranges are included in column 6 of Table 2. 
 
\subsection{Radio 20cm data}\label{radio} 
 
Many of the SCUBA sources from previous surveys have been  
detected at radio wavelengths via  
deep (3$\sigma$$\sim$10-100$\mu$Jy) VLA observations (Smail et al 2000, Ivison  
et al. 1998, 2000, Carilli et al 2001, Bertoldi et al 2001) and faint radio sources have been targeted and detected by SCUBA (Barger et al. 2000, Chapman et al. 2001).  
High resolution radio observations can provide very accurate  
positions for SCUBA sources, as well as yielding  
 independent redshift estimates  
and morphological information.  Using the tight far-IR-radio  
correlation for starburst galaxies (Helou, Soifer \& Rowan-Robinson  
1985, Condon 1992), Carilli $\&$ Yun (1999, 2000) produced a redshift  
indicator based on the predicted redshift dependence of the  
observed spectral index between 20cm and  
850$\mu$m.  Dunne et al. (2000) have produced a similar  
radio-submillimetre redshift indicator based on data from a large  
(104-source) sample of low-redshift galaxies.  Both studies suggest  
that values of $\alpha_{20cm}^{850\mu m} \geq +0.5 $ (where $f_{\nu} \propto  
\nu^{\alpha}$) places  
sources at high redshift (z$\geq$1). 
Assuming that the SCUBA sources detected in this survey have properties 
not dissimilar to the dust-enshrouded galaxies used in the calibration of  
these relations,  
we can expect radio flux densities  $\simeq$0.1~mJy at 20cm.  Medium deep 20cm  
surveys have been performed by Ciliegi et al. (1998) in the  
ELAIS N2 area to a maximum depth of 0.135mJy (5$\sigma$) and  
by de Ruiter et al.(1997) in the Lockman Hole area to a comparable  
depth.   Three sources in the Lockman Hole area (LH850.1, LH850.8  
and LH850.12) have close radio associations.  The remaining sources  
fall below the respective detection limits of the two radio surveys. 
 
All bar one (LH850.8, but see Section \ref{kband} for discussion) of the SCUBA sources have $\alpha_{20cm}^{850\mu m}  
\geq +0.6$.  Employing the results of Carilli $\&$ Yun (1999, 2000)  
and Dunne et al. (2000) leads to the conclusion that, once again, virtually all 
these sources (ie 18/19) must lie at redshifts greater than $\simeq$1.0 
(based on the mean values of $\alpha_{20cm}^{850\mu m}$  
(see Table \ref{redshifts}).  The mean redshift limit for the sample based  
on this $\alpha_{20cm}^{850\mu m}$ indicator is  
$\langle z_{lim} \rangle = 1.5$.   
 
Deeper radio observations of both fields are currently underway, and  
will be reported by Ivison et al. (2001). 
 
\subsection{175/850$\mu $m flux-density ratio} 
 
When the 8-mJy survey was first designed it was anticipated that the ISOPHOT  
175$\mu m$ surveys of the two selected fields would yield a typical  
3-$\sigma$ flux-density limit for undetected sources of $S_{175\mu m} < 50$mJy. 
Such limits would be of considerable interest because, for a reasonable  
range of assumed SEDs, a non-detection of an 8mJy SCUBA source at this 175$\mu m$  
level would imply $z > 2$. Unfortunately, in practice the ISOPHOT surveys  
have not come close to achieving their originally-predicted sensitivities  
(failing by a factor  
of at least $\simeq 3$) and consequently 
the actual redshift constraints provided by the ISOPHOT 
coverage of our survey fields are generally  
weaker than those already derived above 
from the 450/850 and 20cm/850$\mu m$ flux-density ratios.  
However, for completeness we 
note that the non-detection of all the ELAIS N2 sources in the 175$\mu m$  
ELAIS survey ($S_{175\mu m} < 150$ mJy) does still imply a  
minimum redshift of $z > 1$ for all the  
SCUBA sources. This limit is included in column 5 of  
Table 2 as it does at least represent one 
further piece of independent evidence in support of the basic case that  
essentially all the bright SCUBA sources uncovered in the 8mJy survey  
lie at $z > 1$. 
 
\section{Candidate optical, near-infrared and X-ray counterparts} 
 
\subsection{Optical data}\label{rband} 
 
Deep $R$-band and $I$-band images of the ELAIS N2 area were taken  
using the PFC on the WHT during 1999 and 2000 (Willott et al. 2001, in prep),  
both covering 0.07 sq. degrees reaching limiting depths of $R=27$  
and $I=26$ (measured through a 1.5-arcsec radius aperture).   
An $I$-band image of the Lockman Hole E area has also  
been recently obtained with the PFC on the WHT (Ivison et al. 2001), 
this time reaching a limiting depth of $I \simeq 24.5$.   
A deeper $I$-band image (reaching $I=26$ through a 4-arcsec aperture)  
has also been obtained for LH850.1 and LH850.8.  
 
Postage stamp images have been extracted from these optical images, covering  
$30^{\prime \prime} \times 30^{\prime \prime}$ centred on the position of each SCUBA source. 
These images are shown in Figures 4 and 5 (Lockman Hole), and in Figure 6 
(ELAIS N2)  
with the positional uncertainty for each SCUBA source indicated by a circle 
of radius 6 arcsec. 
 
It is evident from these postage-stamp images that while for some 
SCUBA sources there exist no potential optical counterparts to the limit  
of these data, in most cases several alternative identifications 
lie within the SCUBA positional error circle. 
The positions of all optical sources within 6 arcsec of each 850$\mu$m  
centroid are listed in Tables 3 and 4, with the corresponding  
aperture magnitudes 
(and in the case of empty fields, limiting magnitudes) 
included in Table 1. 
 
The ambiguity surrounding the correct optical identification for most of 
the SCUBA sources is a result of the substantial uncertainty in the  
position of the 850$\mu m$ source, coupled with the large surface  
density of faint galaxies at the limiting magnitude  
of our deep optical data. To test whether any of these potential  
identifications is statistically compelling we have calculated,  
for every candidate object, the probability 
that a galaxy with the observed optical magnitude (or brighter) could lie 
so close to the SCUBA position by chance. The resulting probabilities 
($P_E$; see Downes et al. 1986) 
are given for every candidate optical identification in Tables 3 and 4. 
We stress that these probabilities are often substantially higher 
than the raw Poisson probabilities (Downes et al. 1986).  
This is because the large search radius coupled  
with the high surface density of  
faint optical galaxies means that the vast majority of the SCUBA 
sources have at least one potential optical counterpart. 
 
Unsurprisingly, the values of $P_E$ derived for all but one the potential 
optical identifications are not, at this stage, compellingly 
small ($P_E < 0.05$). This result in part re-affirms the importance  
of future deeper radio and mm interferometric observations for  
reducing the search radius for potential  
counterparts. However, the current calculations 
are still of importance because they quantify the fact that at most  
one of the SCUBA sources (N2850.1) can be statistically associated with 
even a moderately bright optical counterpart. This result in itself provides 
further (completely independent)  
support for the conclusion arrived at above on the basis of SED  
constraints, that essentially all the SCUBA sources uncovered by the 8-mJy  
survey lie at $z > 1$. 
 
It is interesting to consider in more detail the significance 
of the one statistically convincing optical identification uncovered by this  
analysis, namely that provided by both the $I$-band and $R$-band imaging 
of N2850.1. As can be seen from Table 4 and Figure 6, the $R$-band image 
of this object provides 3 potential counterparts within the SCUBA position  
error circle. The brightest of these lies almost exactly on top 
of the nominal 850$\mu m$ position, while the two fainter options lie 
right at the edge of the adopted search region. Consequently the brightest 
candidate has a very low probability of being a chance coincidence 
($P_E = 0.06$), and indeed the probability of this object 
being a chance coincidence in the associated $I$-band image is even smaller. 
($P_E = 0.01$). Thus, unless future interferometric follow-up should show the 
SCUBA position of this source to be seriously in error, it seems  
highly likely that this $R$/$I$-band counterpart is physically  
associated with the 850$\mu m$ source. However, we would caution  
that this does not necessarily in itself guarantee that this is the correct  
identification, a point which is well demonstrated by the follow-up  
observations of the brightest sub-mm source uncovered by Hughes et al. (1998) 
in the SCUBA image of the Hubble Deep Field. HDF850.1 lies sufficiently 
close ($\simeq 1$ arcsec distant) to the elliptical galaxy 3-586.0 that, as  
pointed out by Downes et al. (1998), the probability that this positional coincidence 
should occur by chance is $P_E = 0.05$, similar to the value  
derived here for N2850.1. However, despite this, subsequently improved 
astrometry provided by IRAM PdB  
and VLA imaging of HDF850.1 (Downes et al. 1999) has not strengthened  
the case for this possible identification. Indeed, based  
on broadband optical-infrared photometry, 3-586.0  
appears to be a very passive elliptical  
at $z \simeq 1.1$, a redshift which is completely  
at odds with that inferred for the SCUBA source  
from SED constraints. Thus, at least in the case 
of HDF850.1 it appears that a low value of $P_E$ has been produced 
not because 3-586.0 is the correct identification, but  
perhaps because it is 
associated in some other way with the SCUBA source, possibly assisting  
its sub-mm detectability via gravitational lensing. It will therefore  
be interesting to see whether further deeper observations of N2850.1   
confirm the apparently convincing identification presented here, or  
reveal a more complex picture analogous to HDF850.1. 
 
Finally, we note that at this stage of the survey  
there are three SCUBA sources associated with {\it optical} blank fields;   
LH850.1 has no candidate identifications to a depth of $I=27.4$, while  
LH850.5 and LH850.14 have no possible optical counterparts to a depth of  
$I=24.5$.  
These 3 objects are therefore particularly  
strong candidates for highly obscured, high-redshift  
galaxies, and in fact LH850.1 has now been discovered (via IRAM PdB 1.2mm  
interferometry combined with very deep UKIRT $K$-band imaging) 
to be a faint and complex ERO at $z \simeq 3$ (Lutz et al. (2001). 
 
This discovery reinforces our confidence  
that (because of our conservative selection criteria)  
the lack of any potential optical counterparts 
for LH850.5 and LH850.14 reflects the nature and/or the  
remoteness of these galaxies, and should not be regarded as  
casting doubt on the reality of the 850$\mu$m sources.

\begin{figure*} 
 \centering 
    \vspace*{20cm} 
    \leavevmode 
\includegraphics{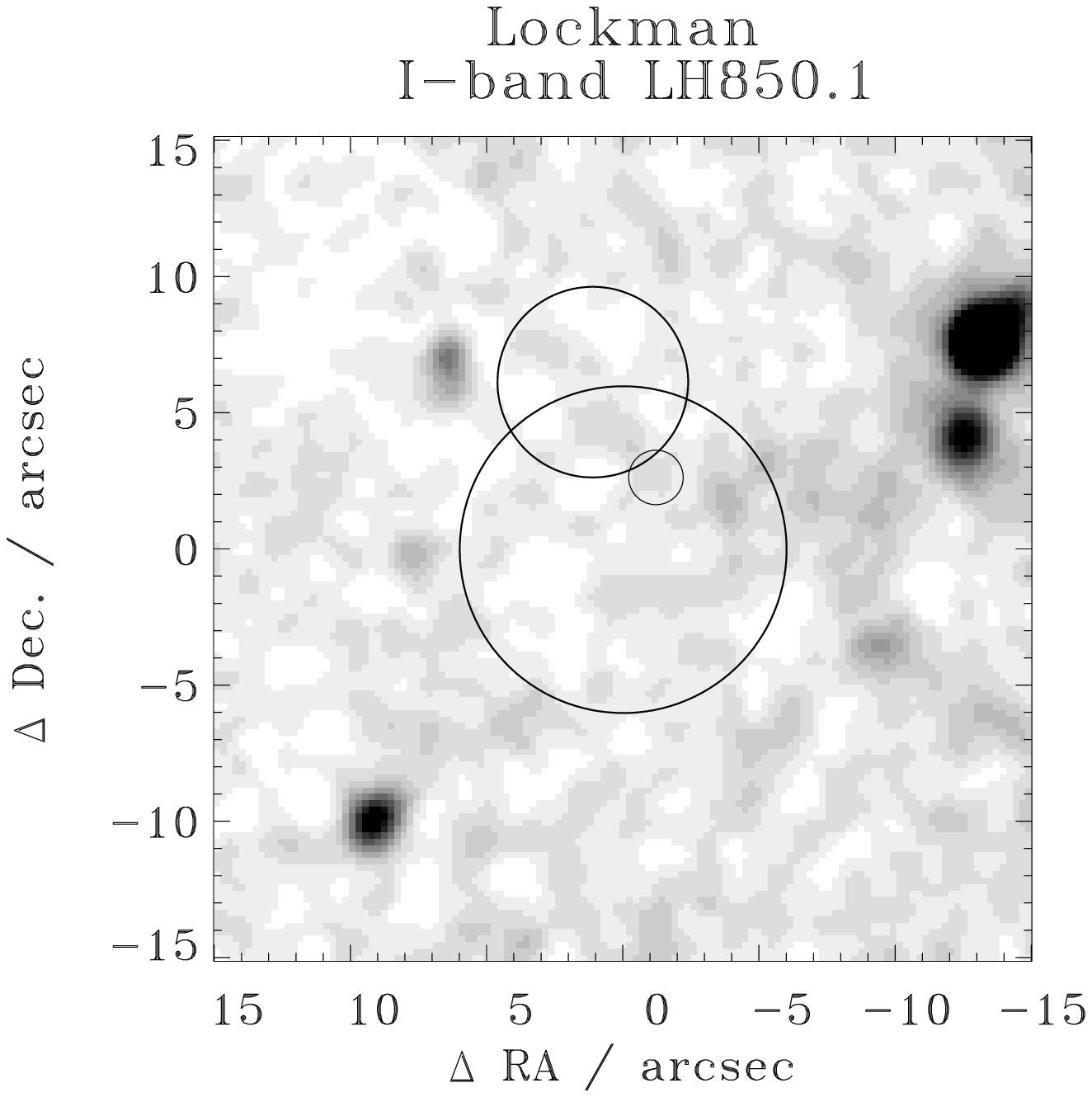}   

 \includegraphics{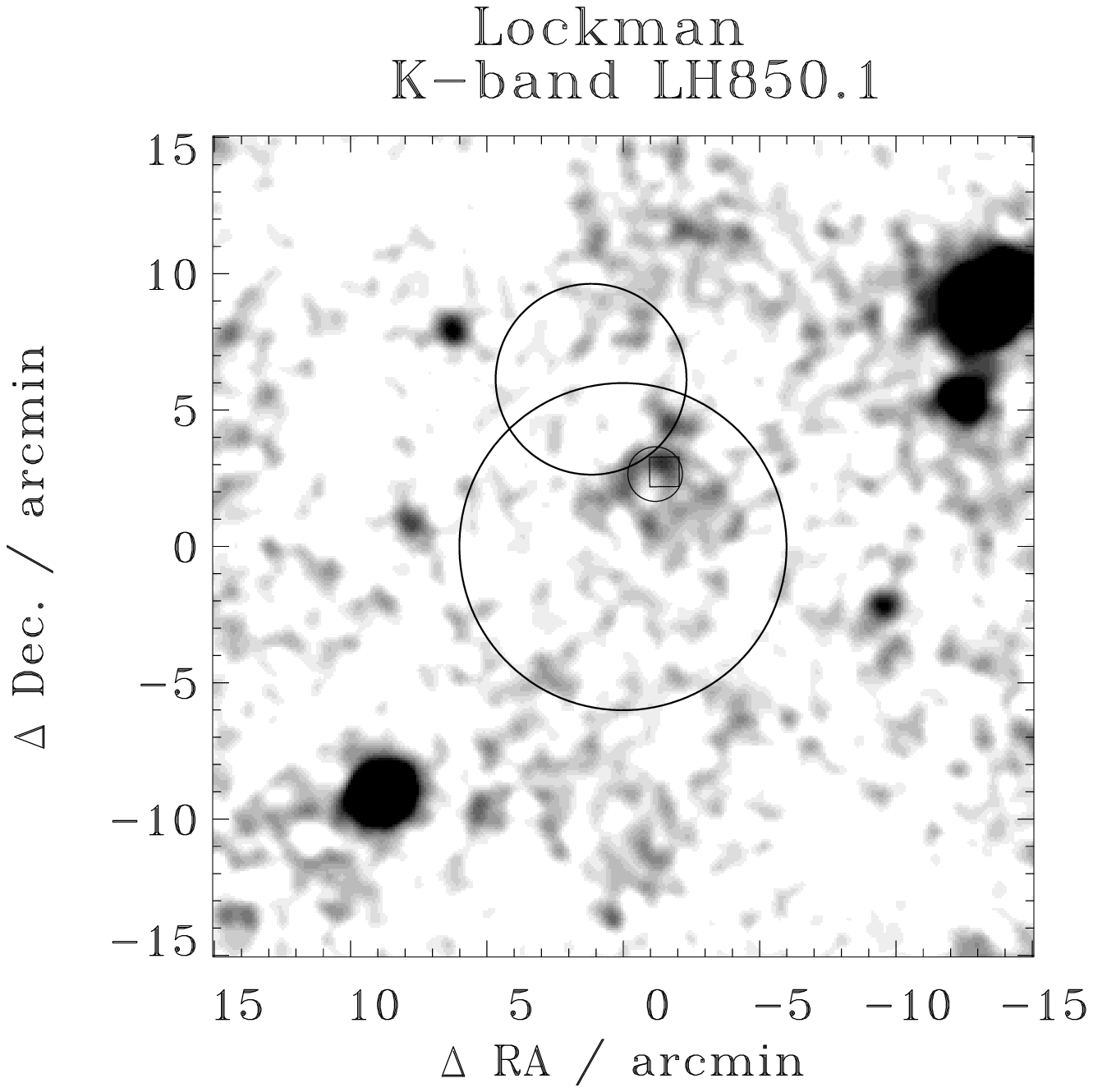}  
 
\includegraphics{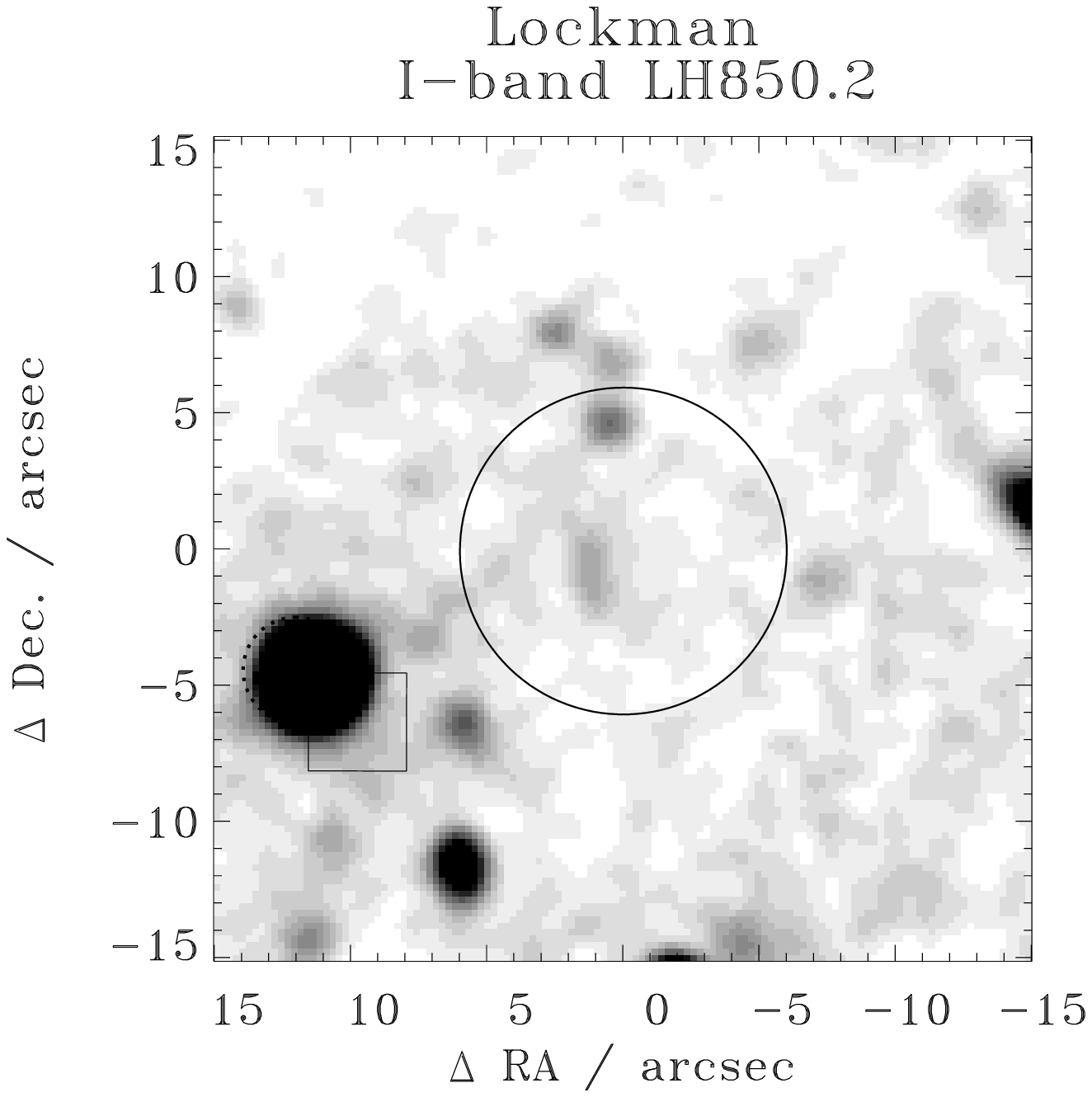}  
 
\includegraphics{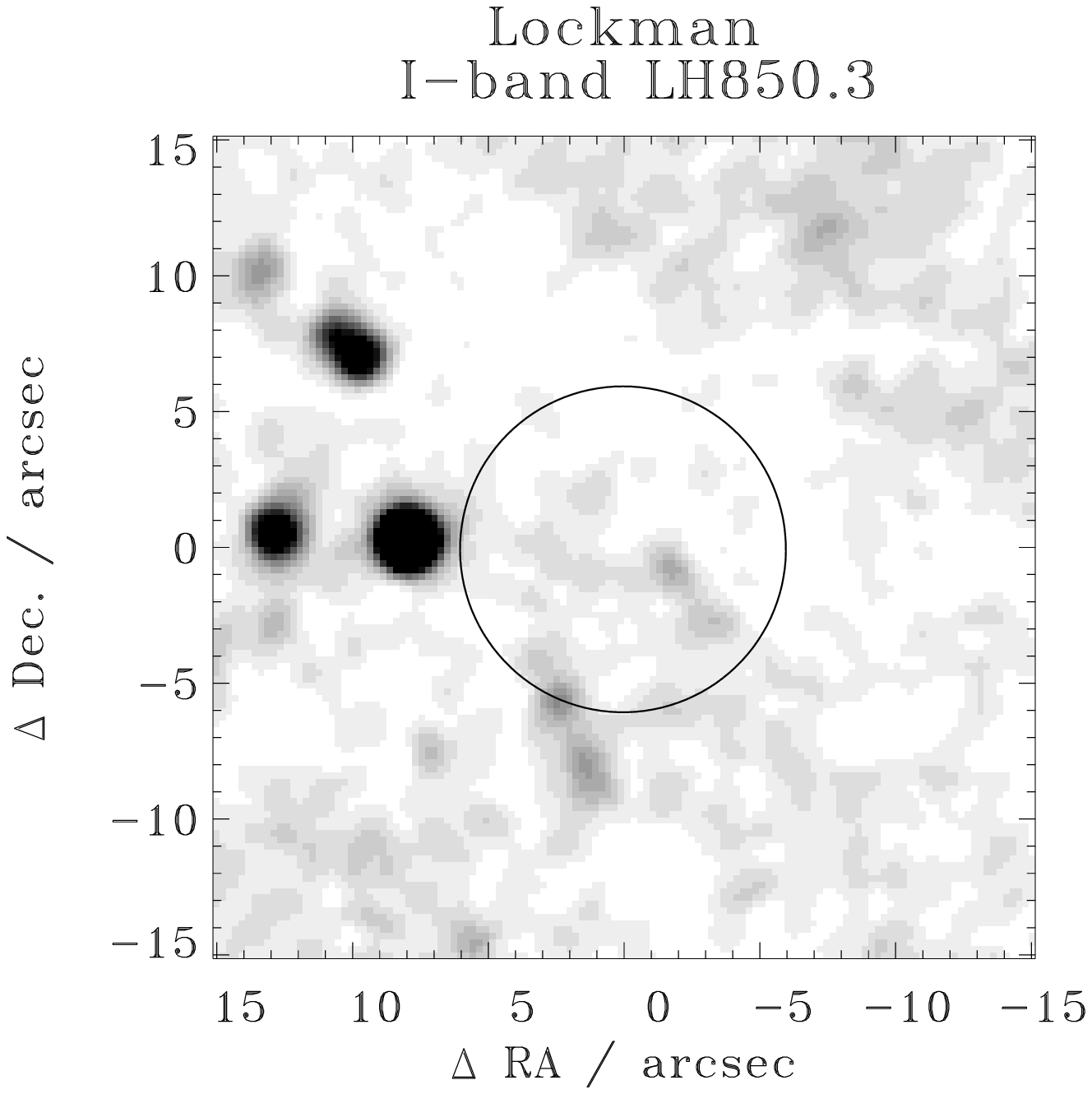}

\includegraphics{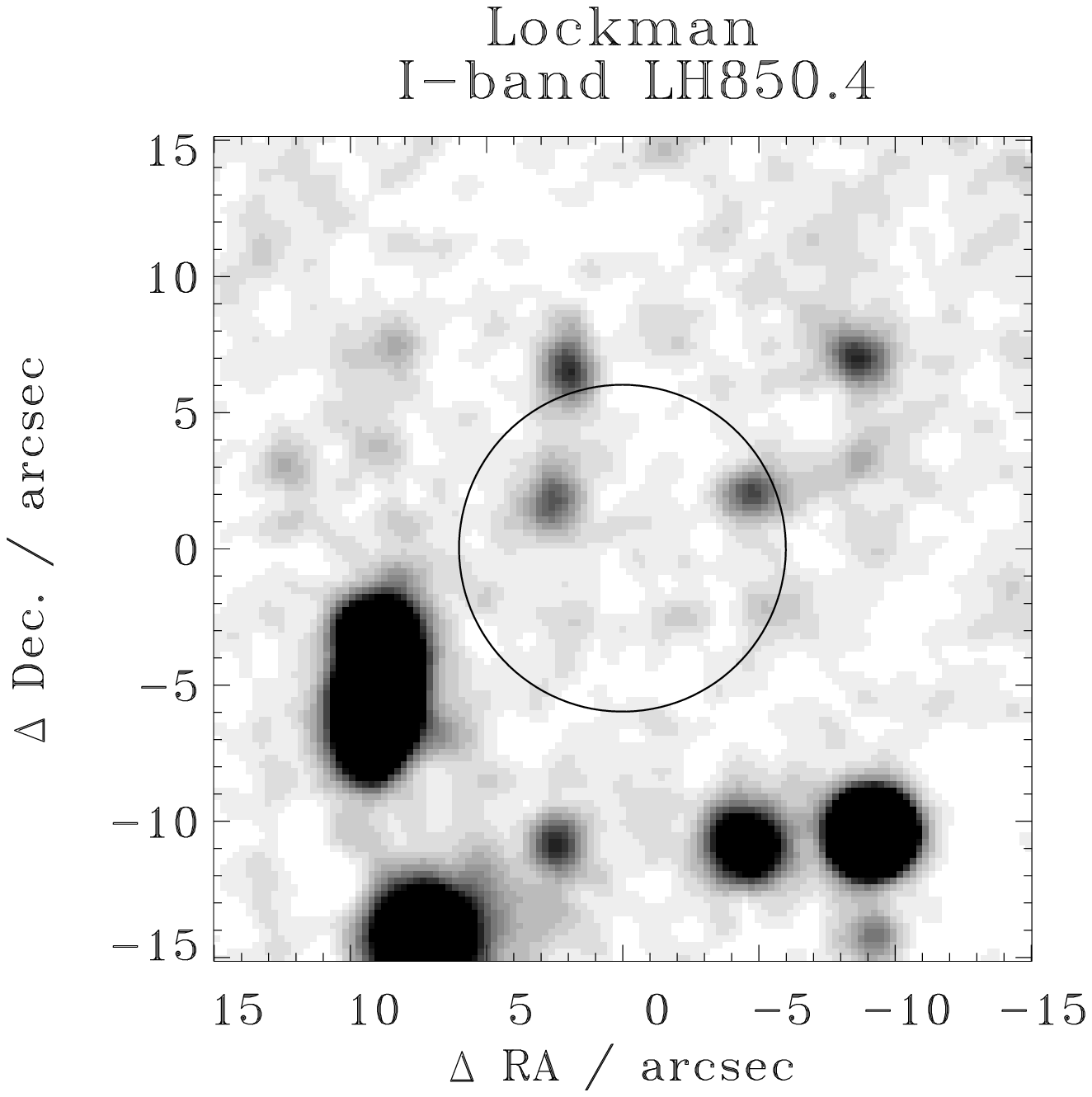} 
\includegraphics{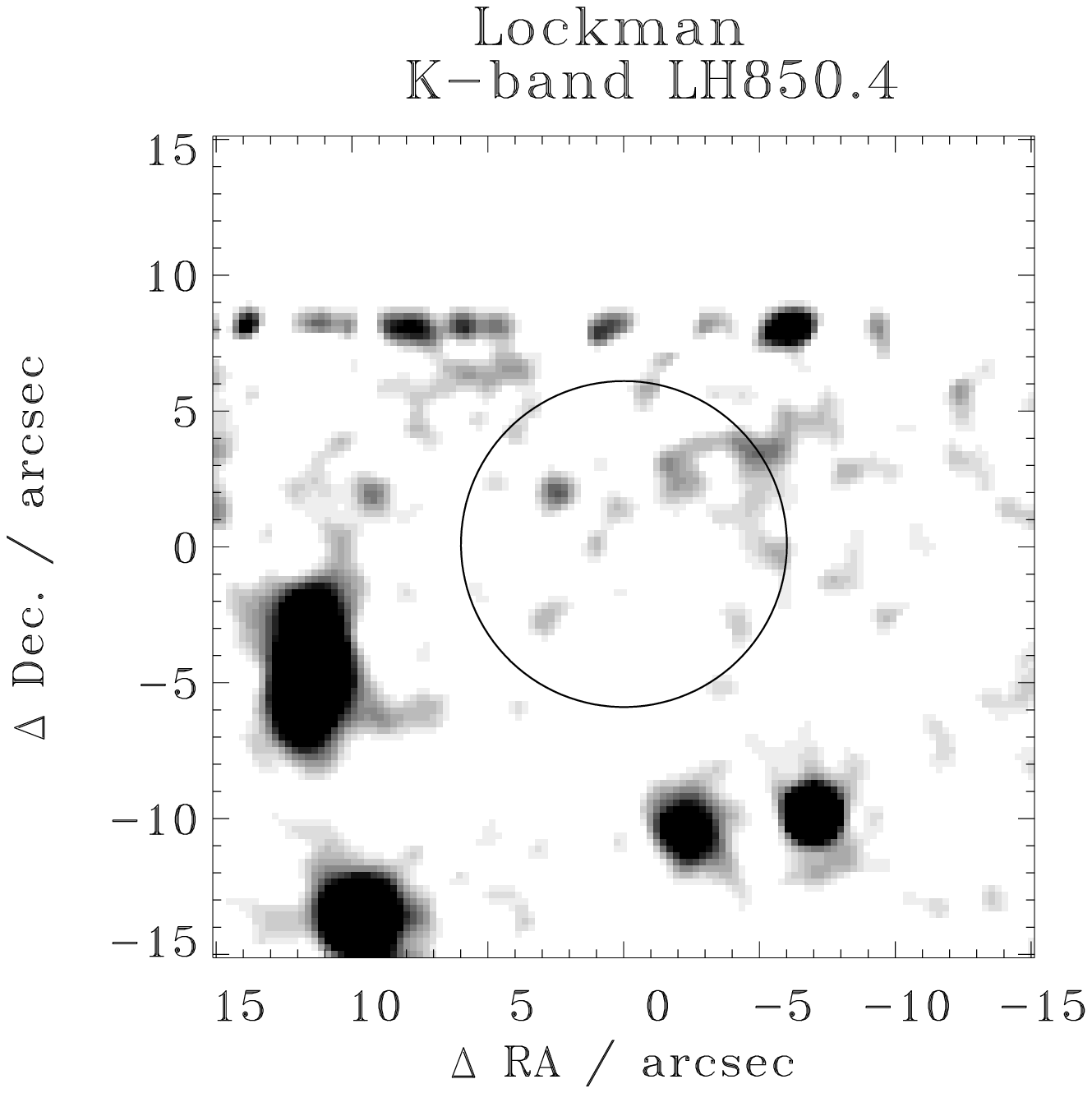}

 \includegraphics{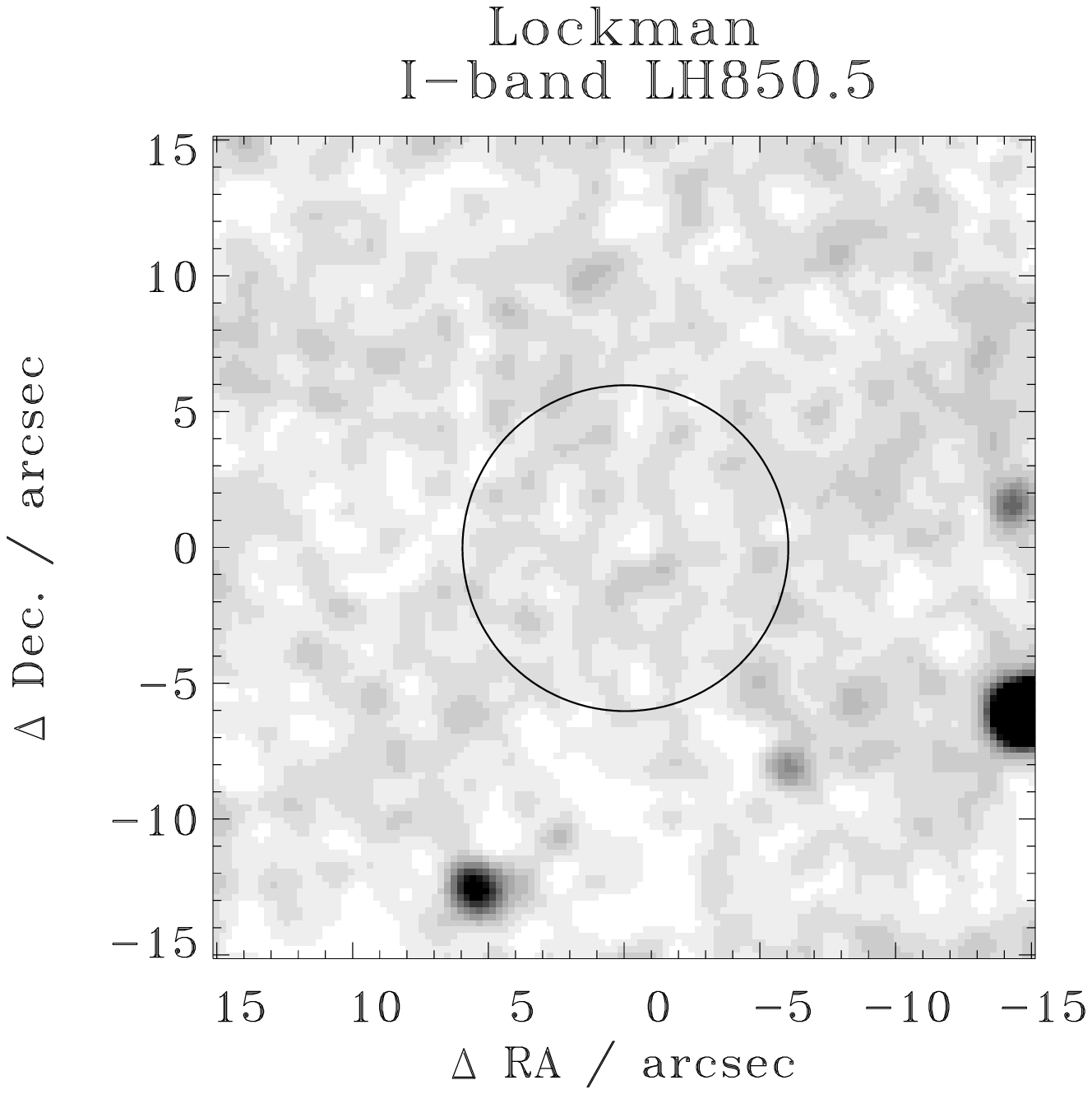} 
\includegraphics{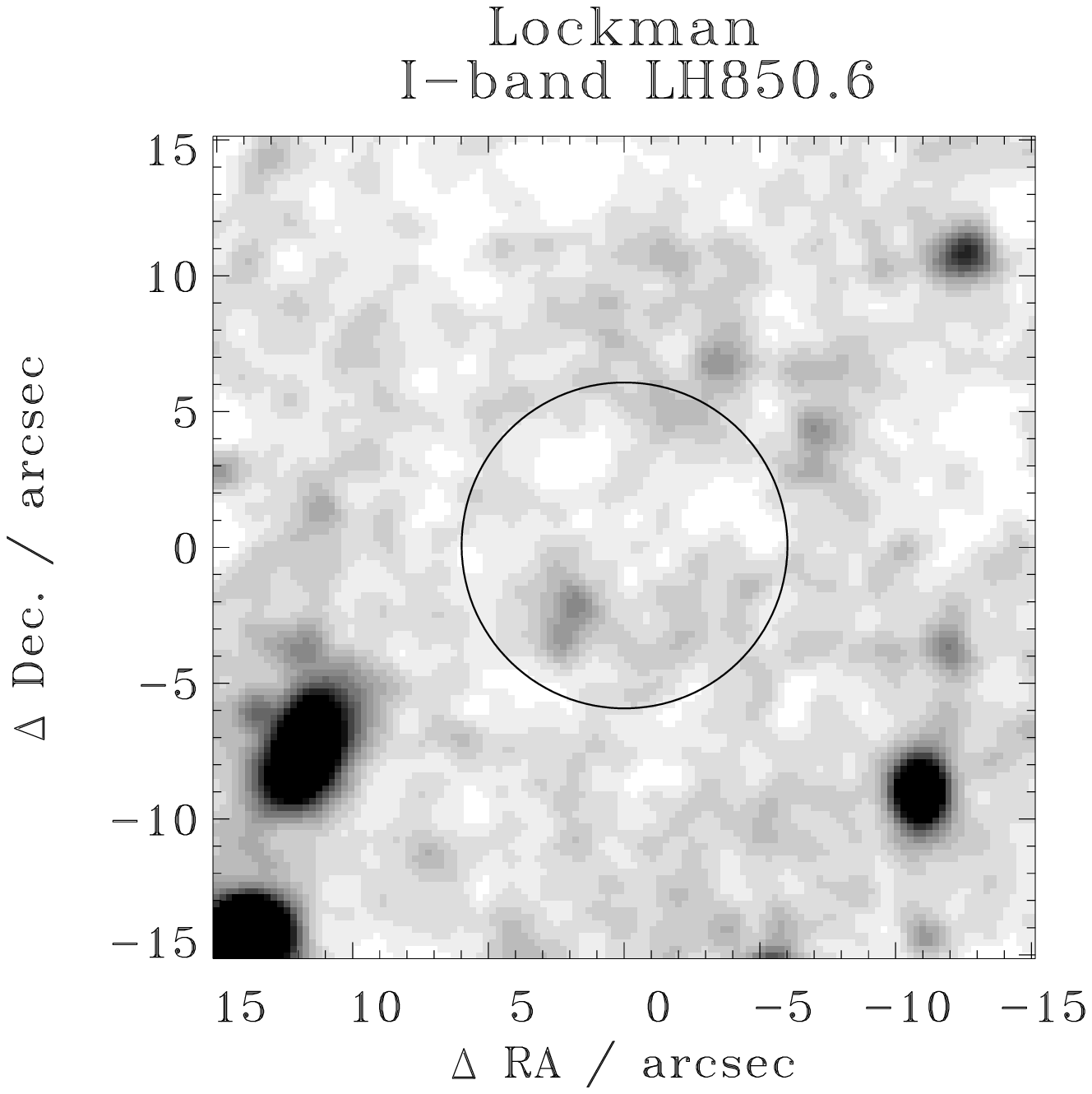}

\includegraphics{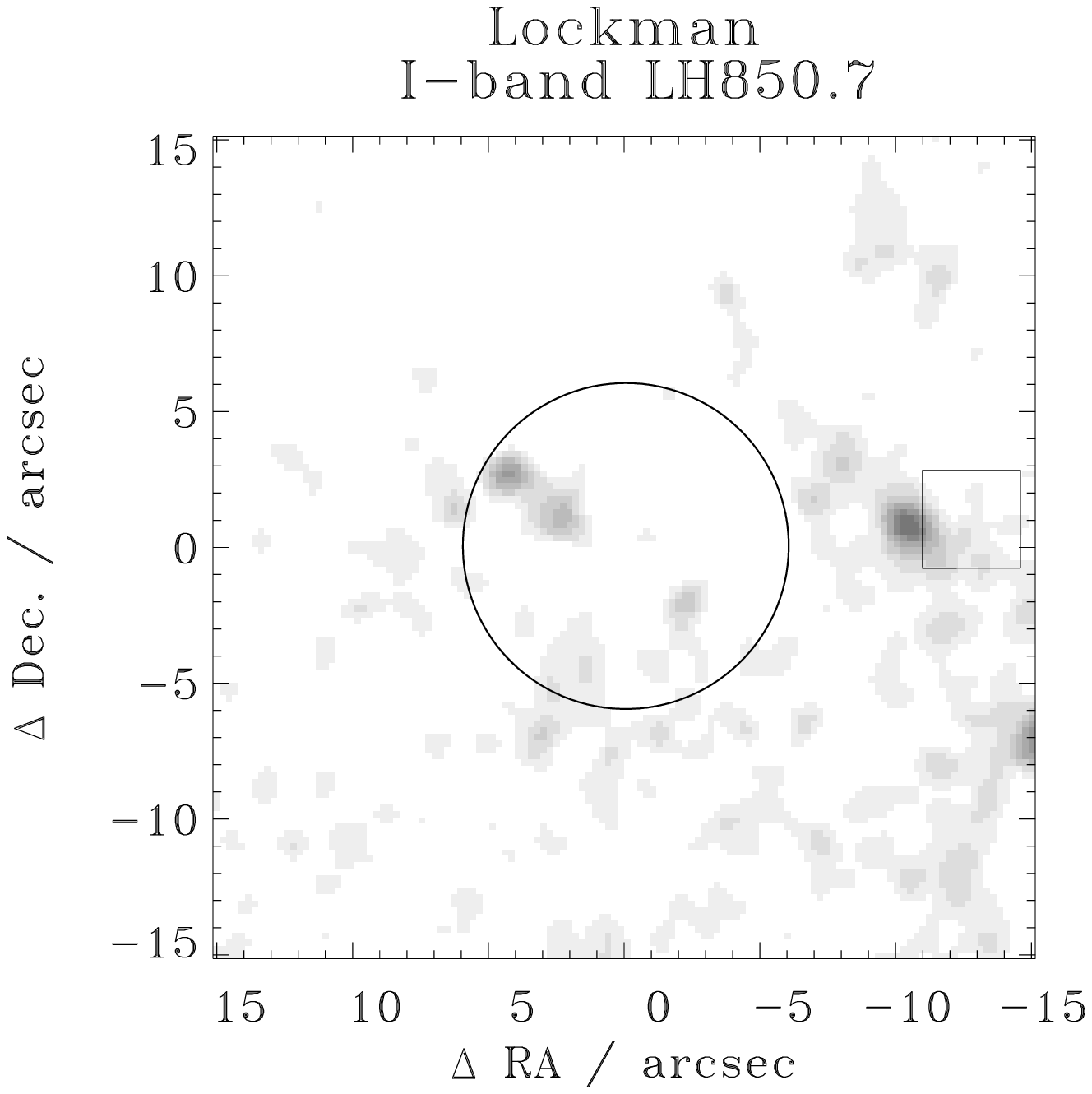}

\includegraphics{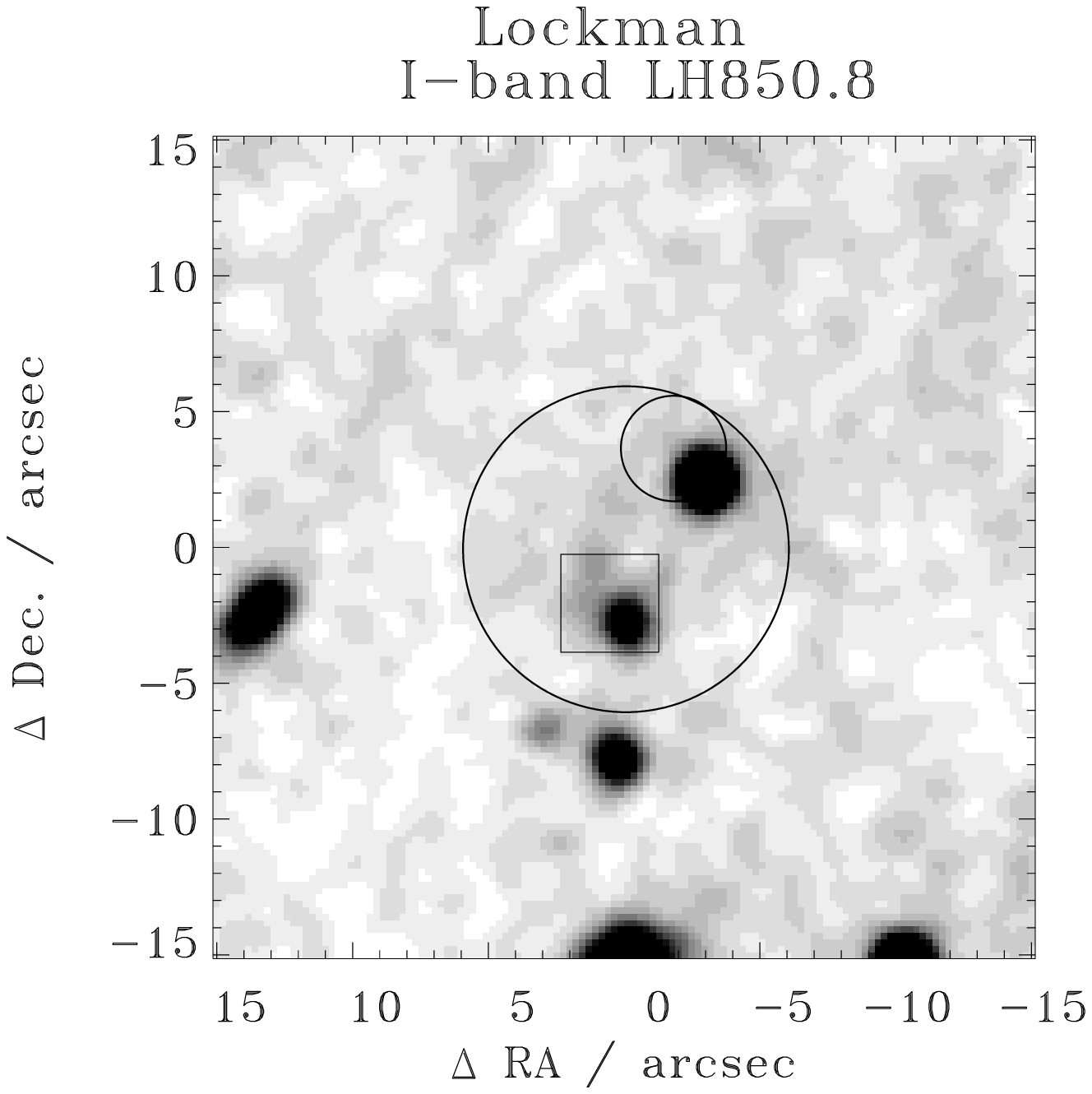}   

 \includegraphics{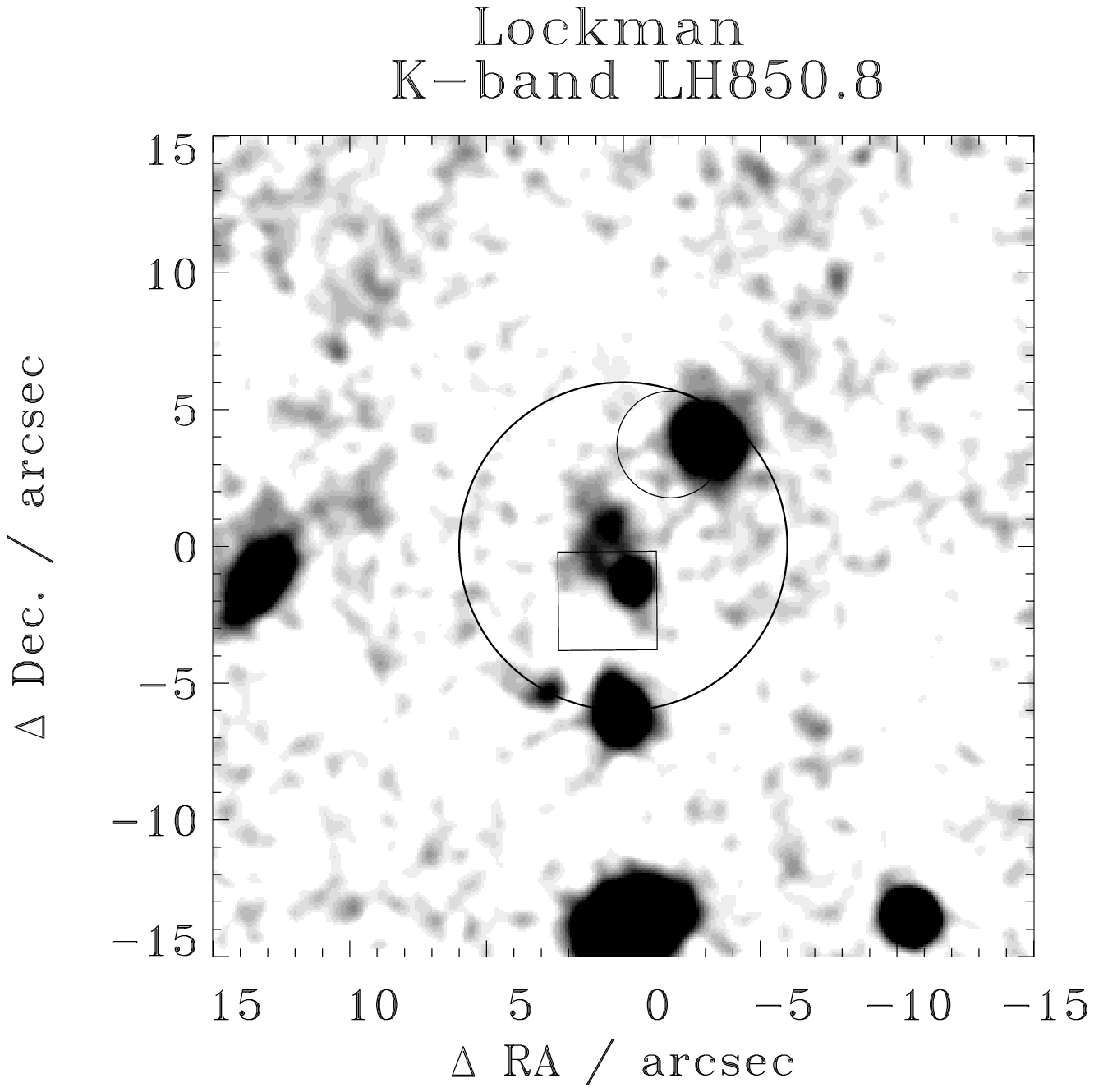}  
 
 \includegraphics{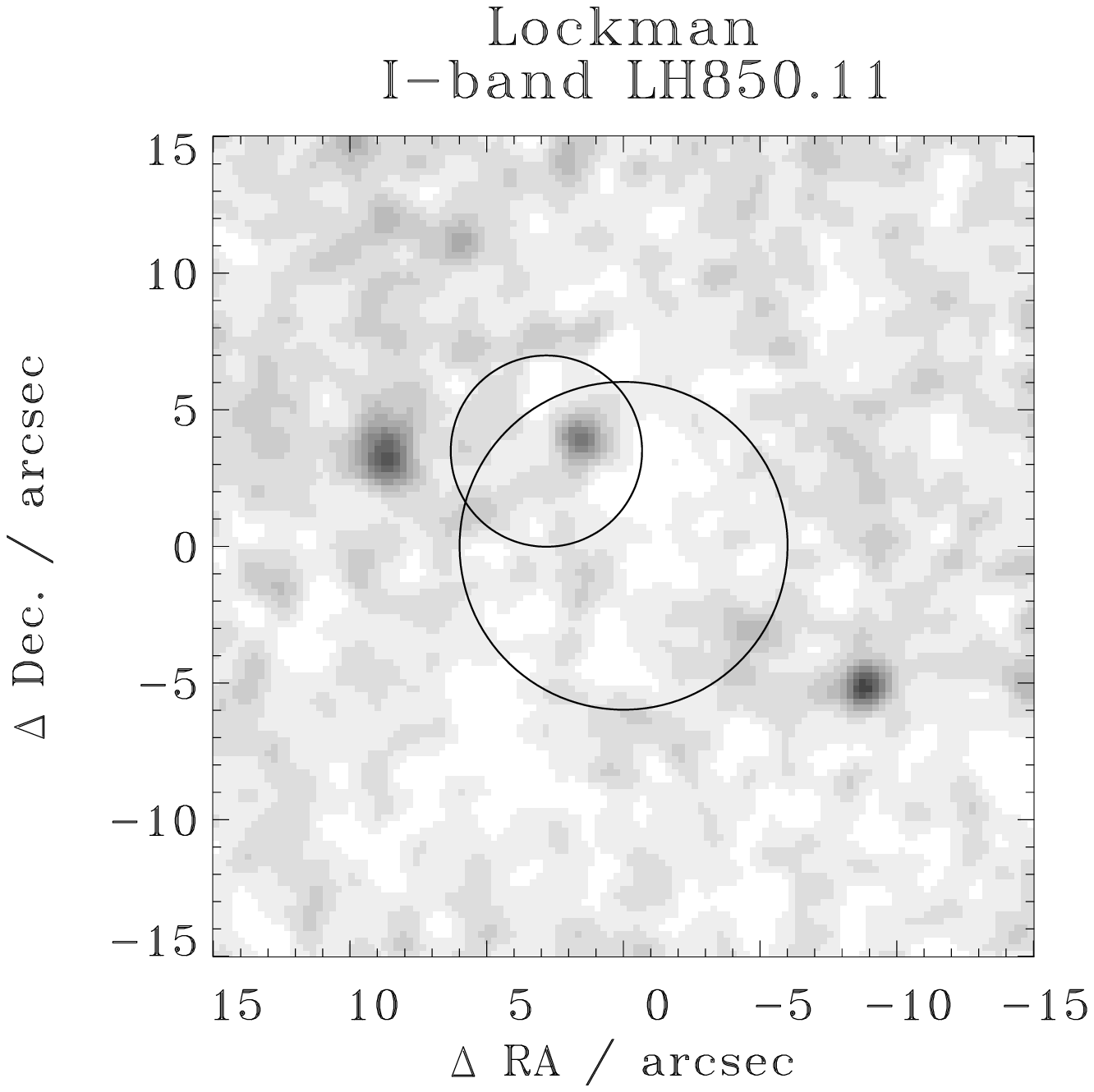} 
\includegraphics{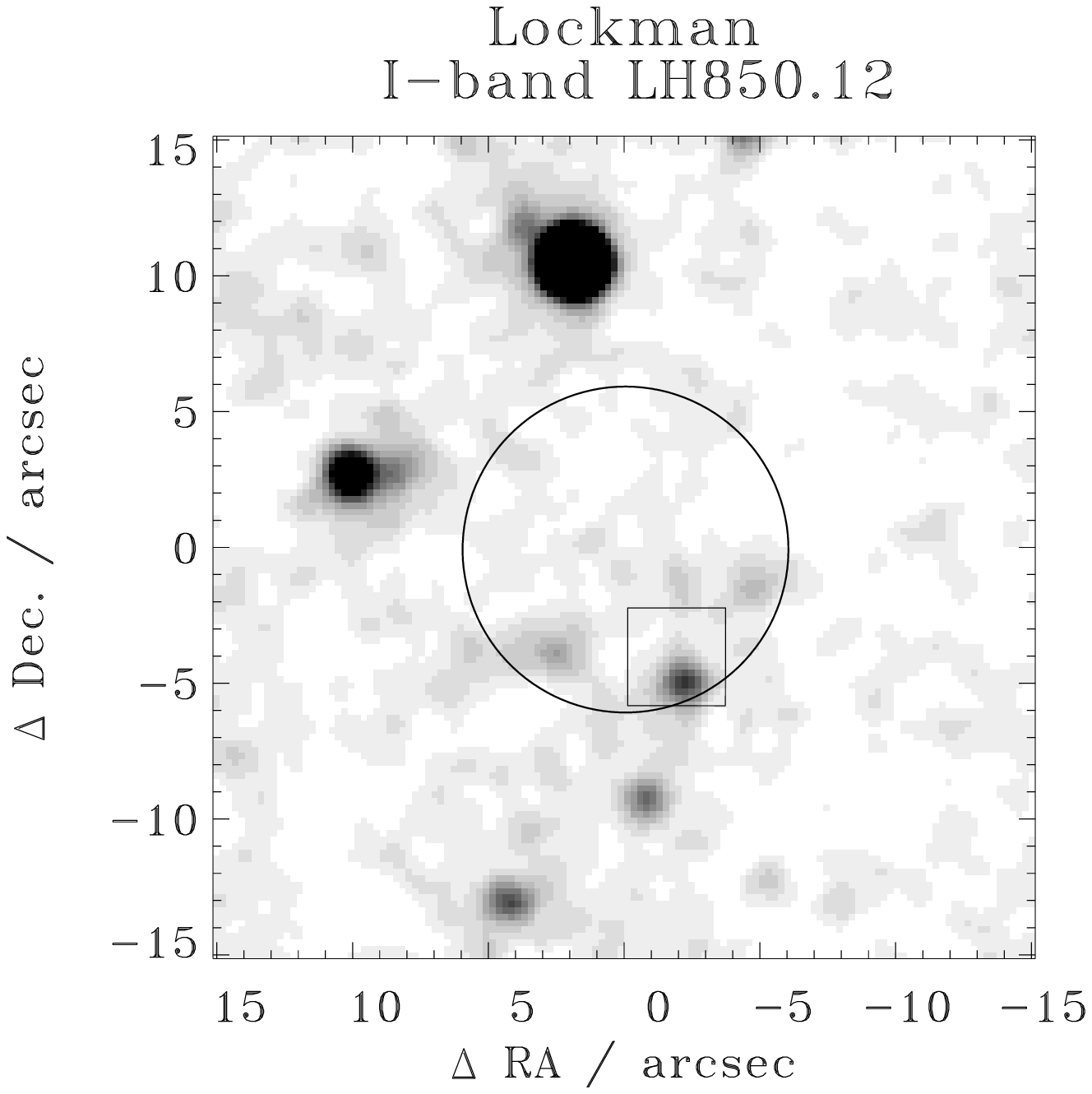}   
\includegraphics{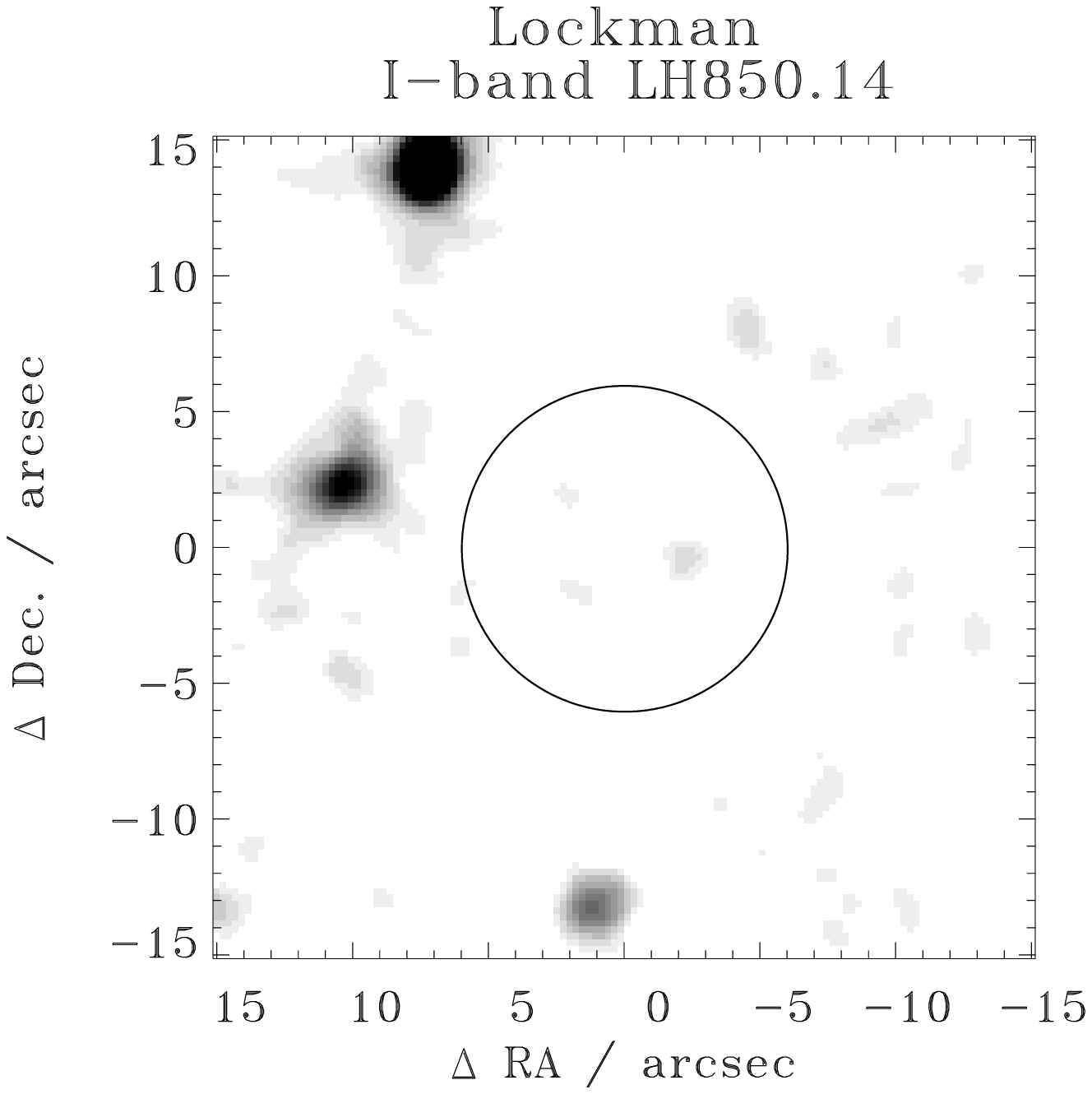} 
 
\includegraphics{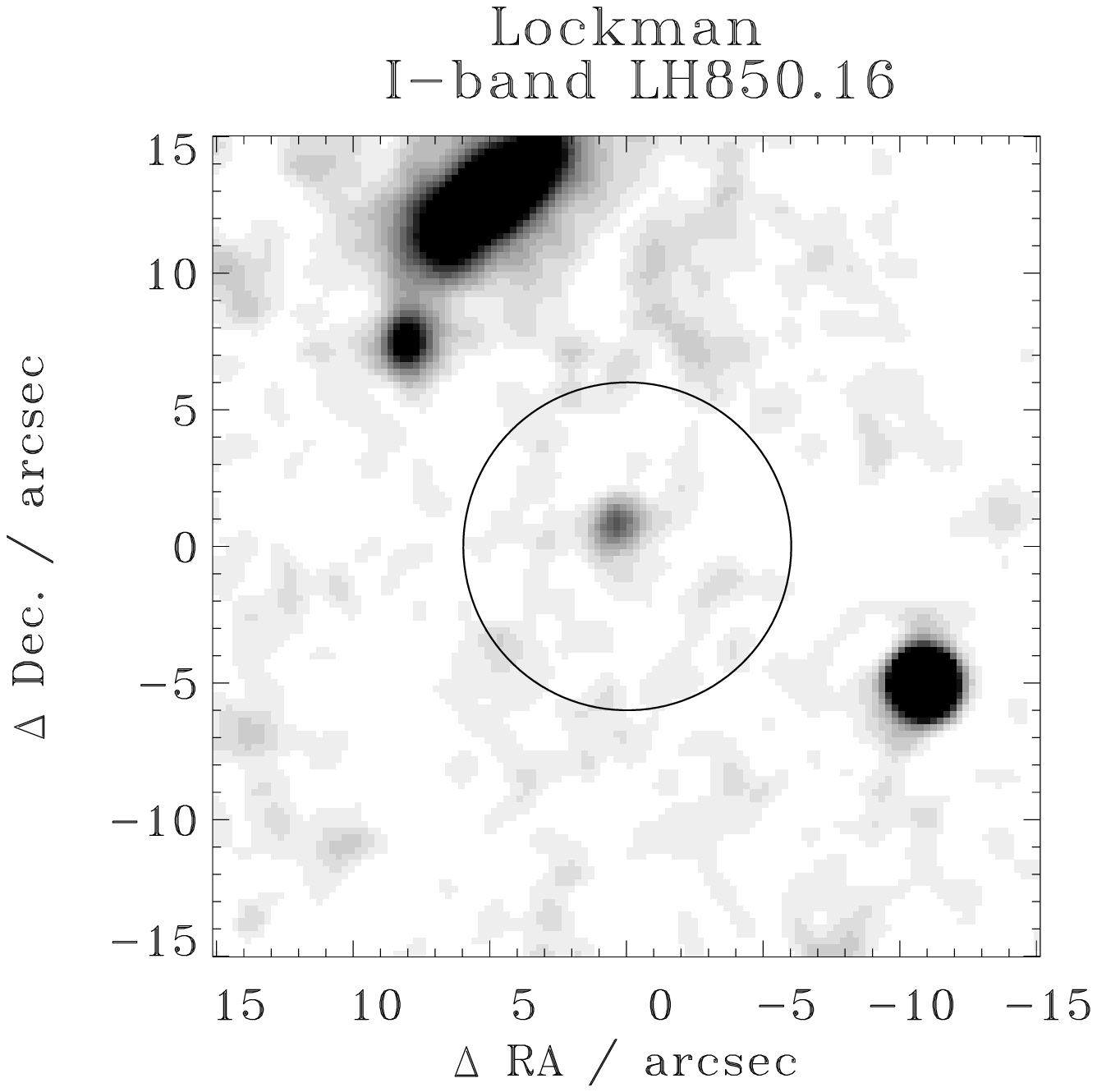}

 \caption[m]{   
$I$ and $K$-band $30 \times 30$ arcsec postage stamps, centred on  
the 850$\mu m$ positions of the Lockman Hole SCUBA sources,   
indicating potential optical and/or near-infrared  
counterparts to the sources uncovered at 850$\mu m$..  
The large circle in each figure has a radius of 6 arcsec, and defines  
the (conservatively large)  
search radius adopted for the calculation of the statistical  
significance of each potential identification as described in section 4.1. 
Top left is LH850.1 which is the most significant 850$\mu m$  
source in our sample, and has been the subject of detailed  
follow-up by Lutz et al. (2001). In this case an additional 
3-arcsec radius circle is included centred on the position  
of our 450$\mu$m detection, with an even smaller circle (1-arcsec radius)  
centred on the position yielded by the  
1.2mm IRAM PdB interferometric detection of Lutz et al. (2001).

The position of the 450$\mu m$ detection of LH850.11 is also 
marked, a position which arguably increases the possibility  
of the single optical counterpart provided by the $I$-band image. 
 
For LH850.7, LH850.8 and LH850.12 boxes have been included  
to indicate the 2-$\sigma$ positional 
uncertainty associated with the nearest radio sources found in the  
VLA survey of de Ruiter et al. (1997).  For LH850.8 the position of  
the X-ray source (denoted as LH850.8(1) in Tables \ref{flux_table}, \ref{redshifts} and \ref{lh_posn}) detected via ROSAT HRI imaging has also 
been marked (small circle). The picture for this object  
is particularly complex/confusing, with the radio  
source being strongly associated 
with one potential optical/IR SCUBA ID, while the X-ray source  
is apparently associated with another. However, neither optical object is in  
itself a statistical compelling counterpart to the 850$\mu m$ source. 
\label{lh_i_postage}} 
 
\end{figure*}

\begin{figure*} 
 \centering 
    \vspace*{4cm} 
    \leavevmode 
 \includegraphics{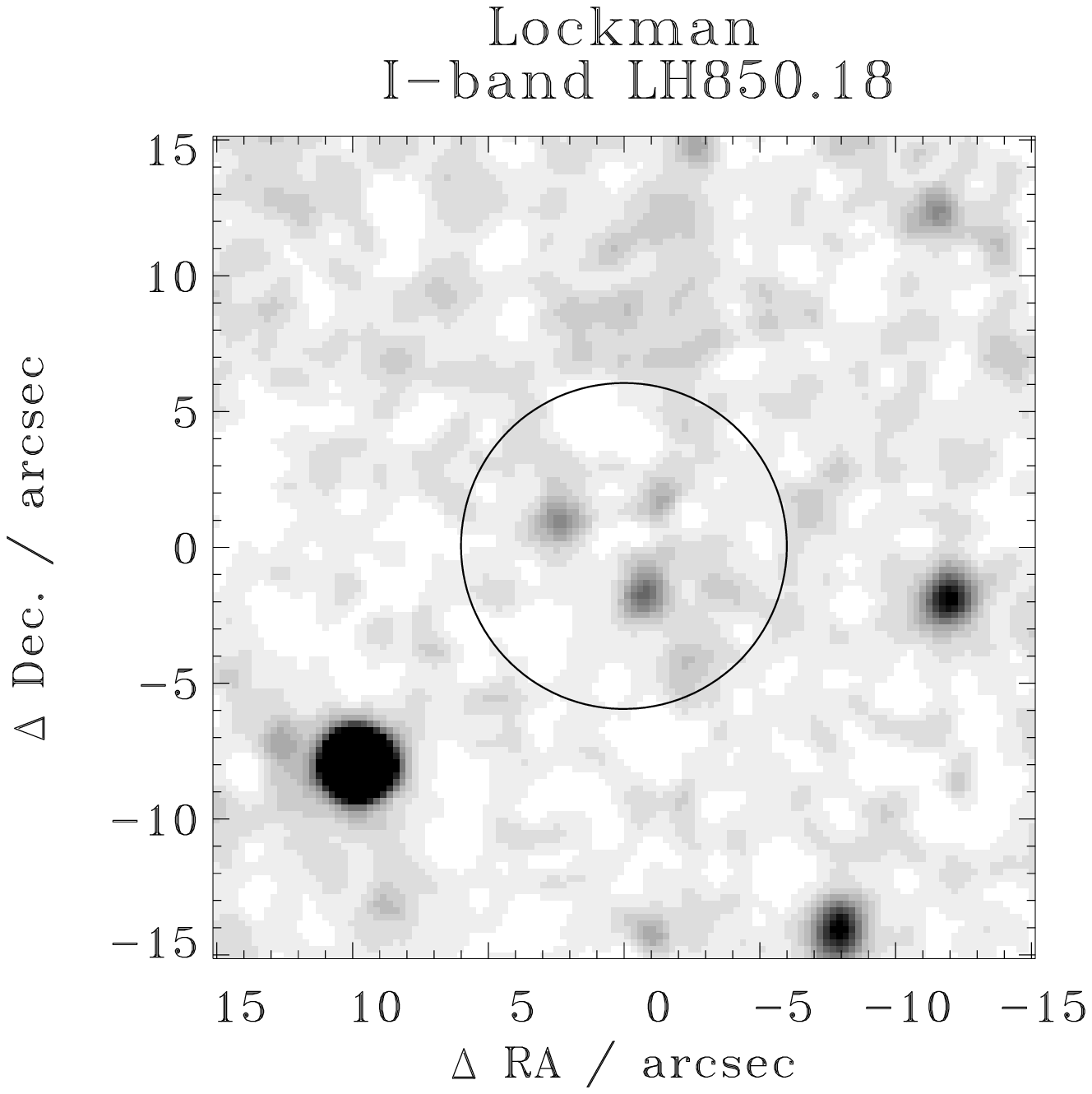}   
\includegraphics{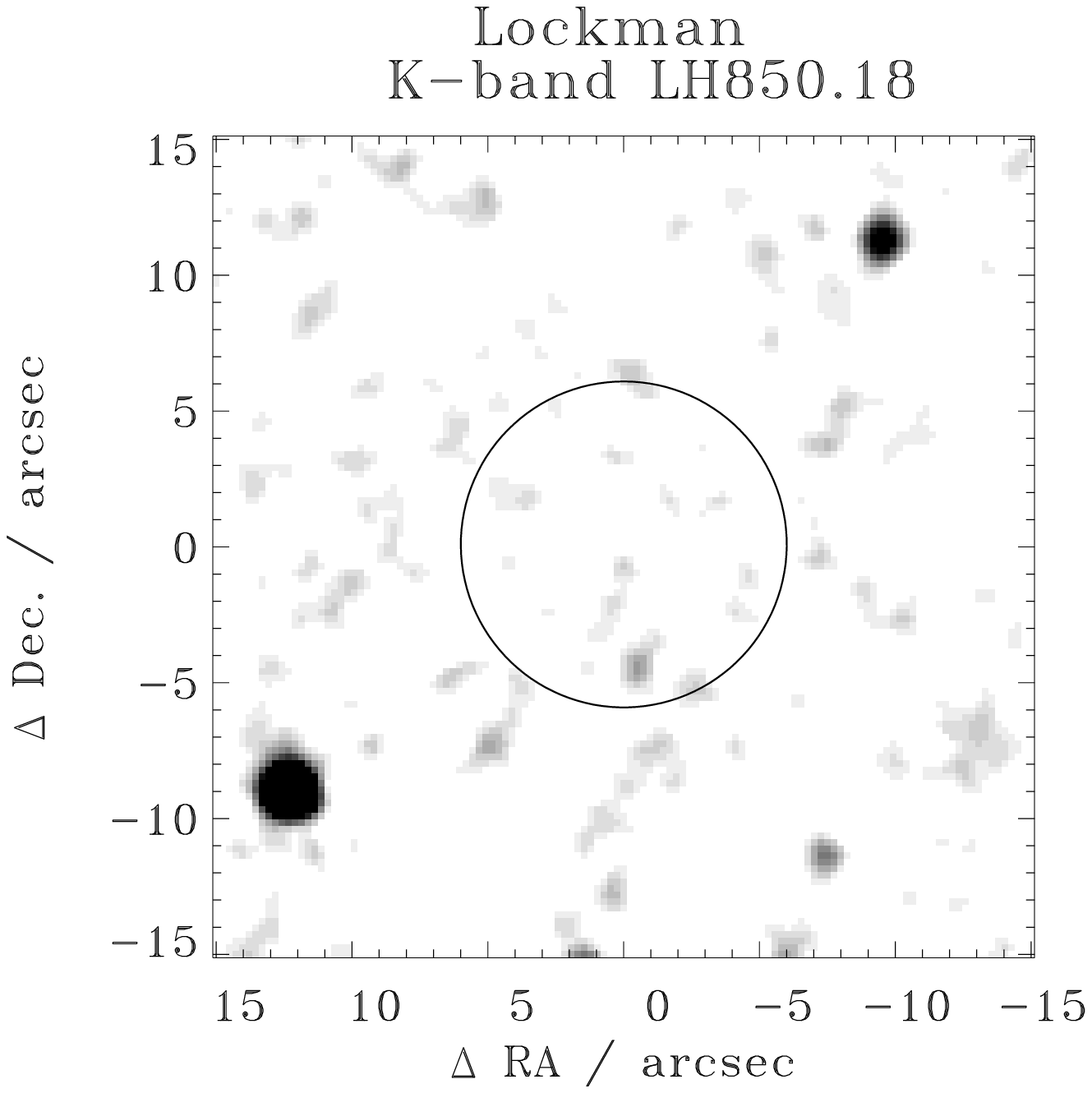} 
 
 \caption[m]{$I$ and $K$-band $30 \times 30$ arcsec postage stamps, centred on  
the 850$\mu m$ position of the Lockman Hole SCUBA source 
LH850.18,   
indicating potential optical and/or near-infrared  
counterparts to the source uncovered at 850$\mu m$.  
The large circle in each figure has a radius of 6 arcsec, and defines  
the   
search radius adopted for the calculation of the statistical  
significance of each potential identification as described in Section 4.1. 
\label{lh_i_postage_2} } 
 
 \end{figure*}

\begin{figure*} 
 \centering 
    \vspace*{12cm} 
    \leavevmode 
 
\includegraphics{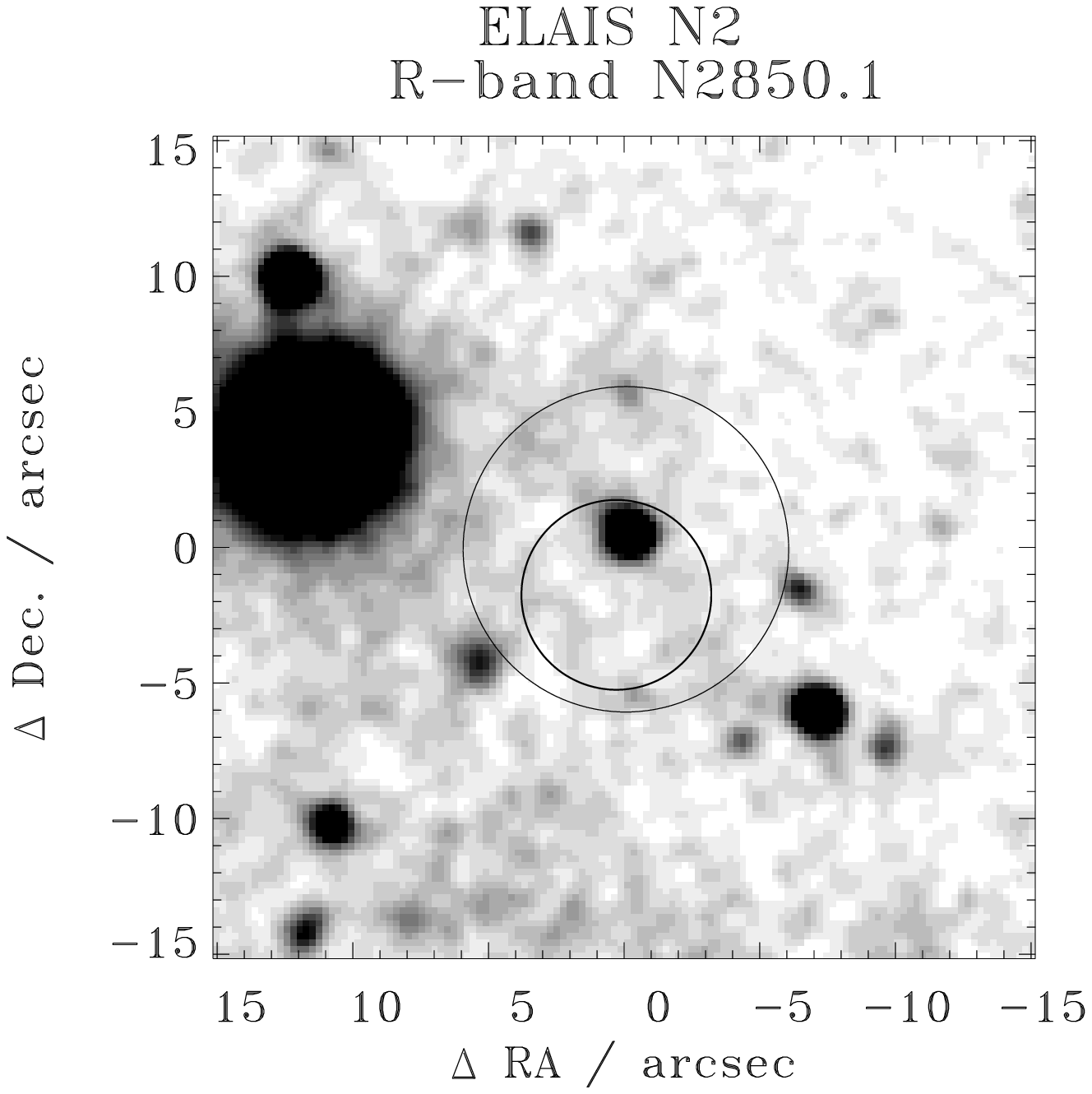}  
\includegraphics{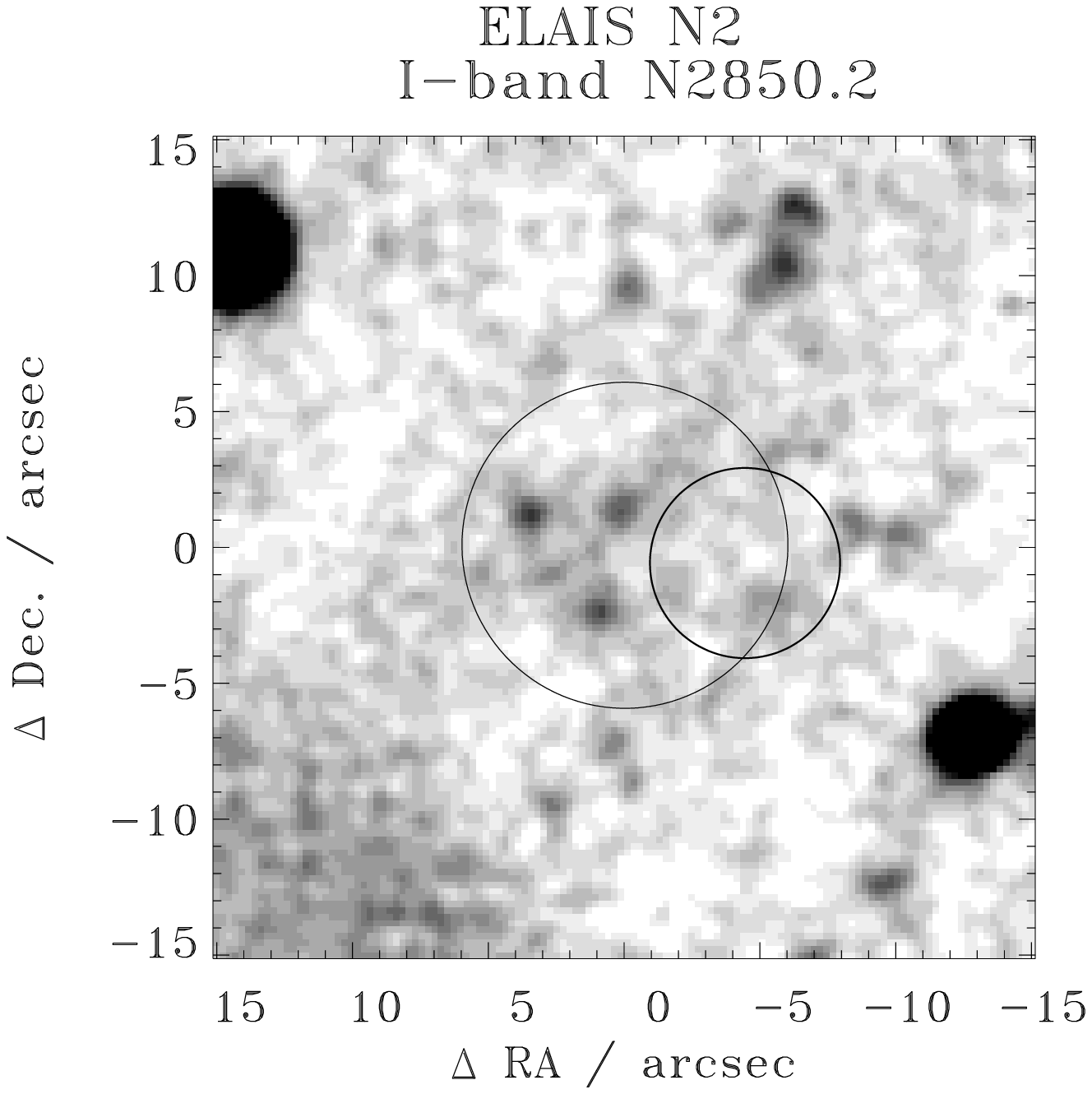}  
\includegraphics{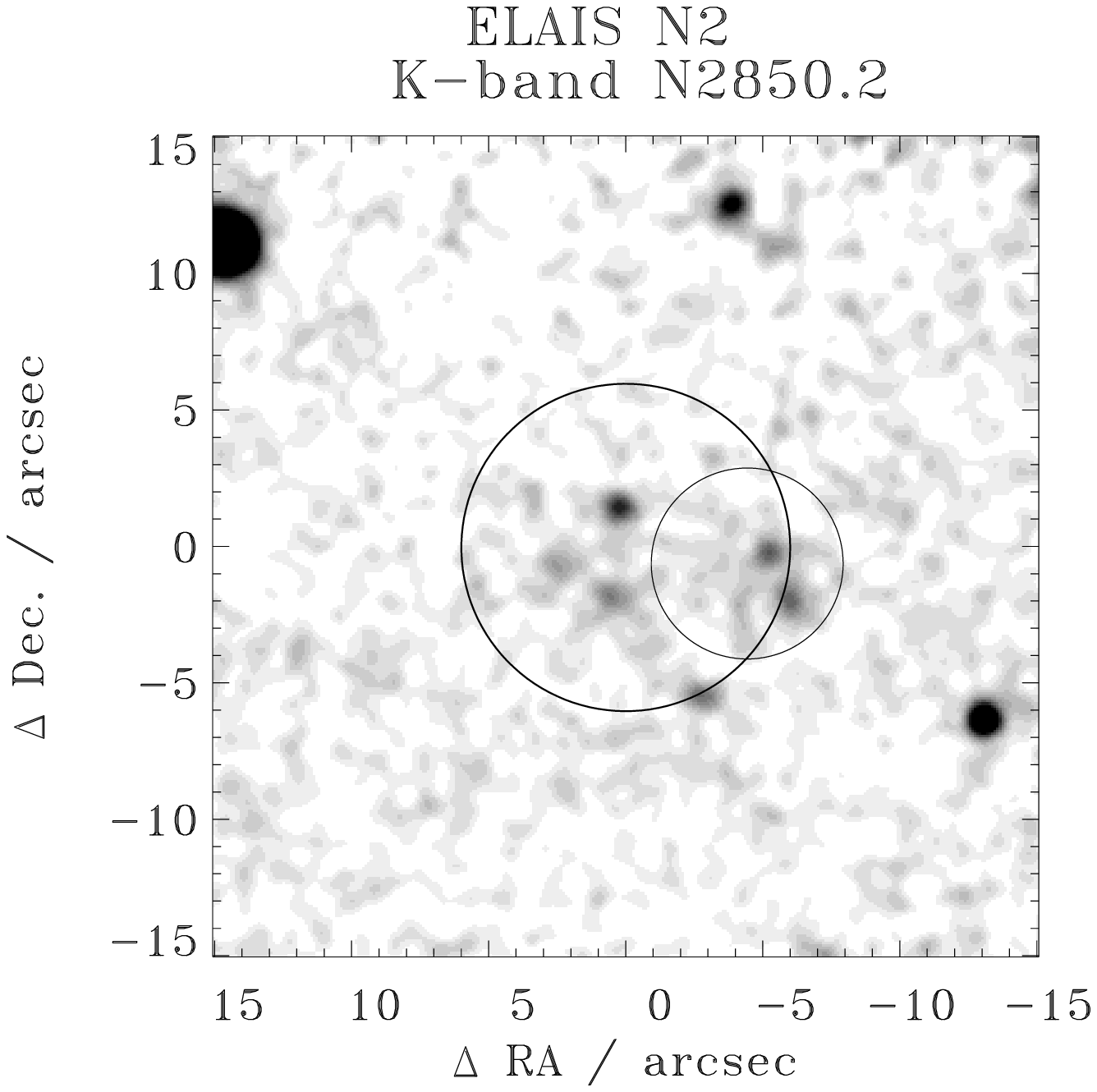} 
 
\includegraphics{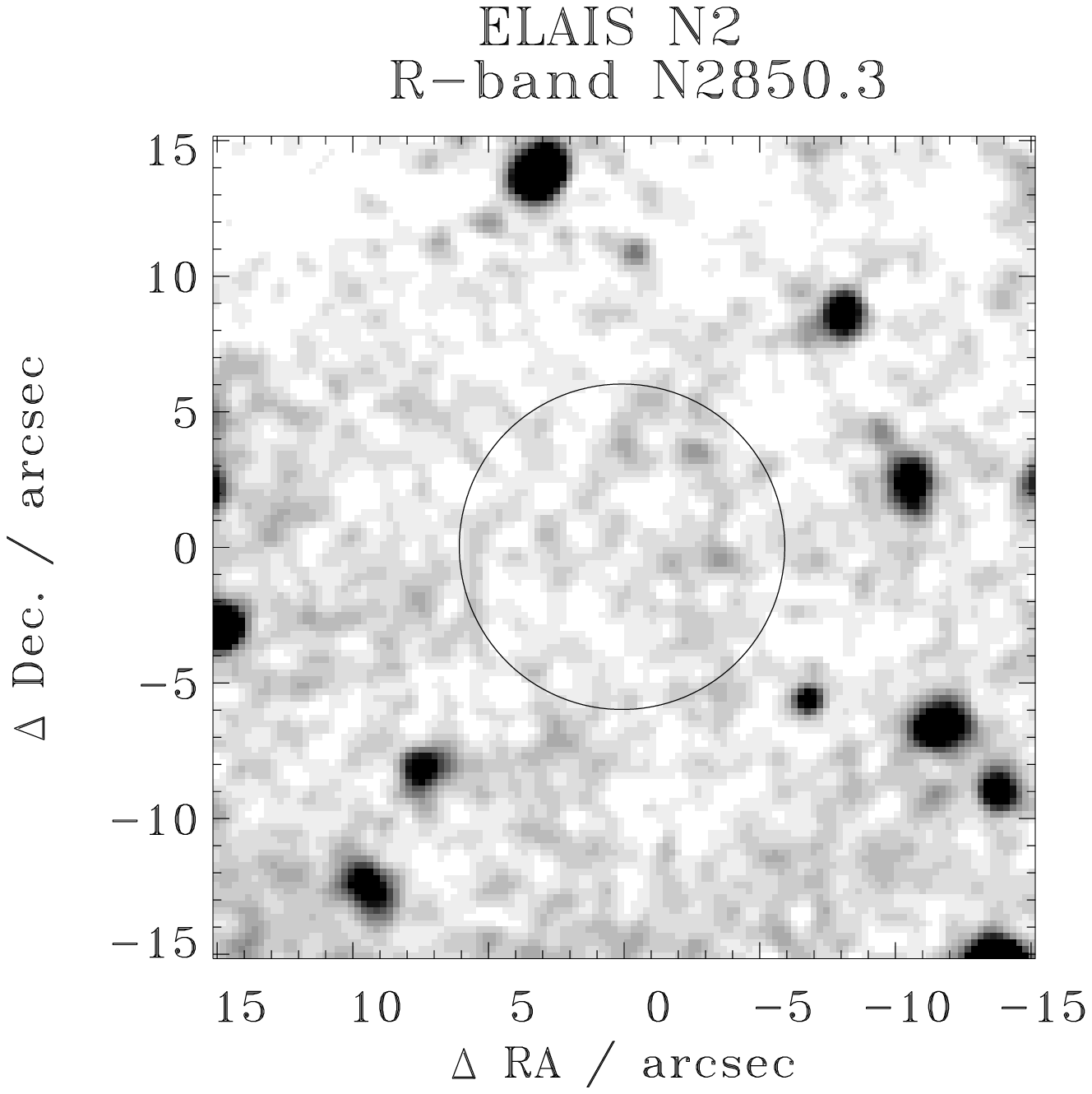} 
\includegraphics{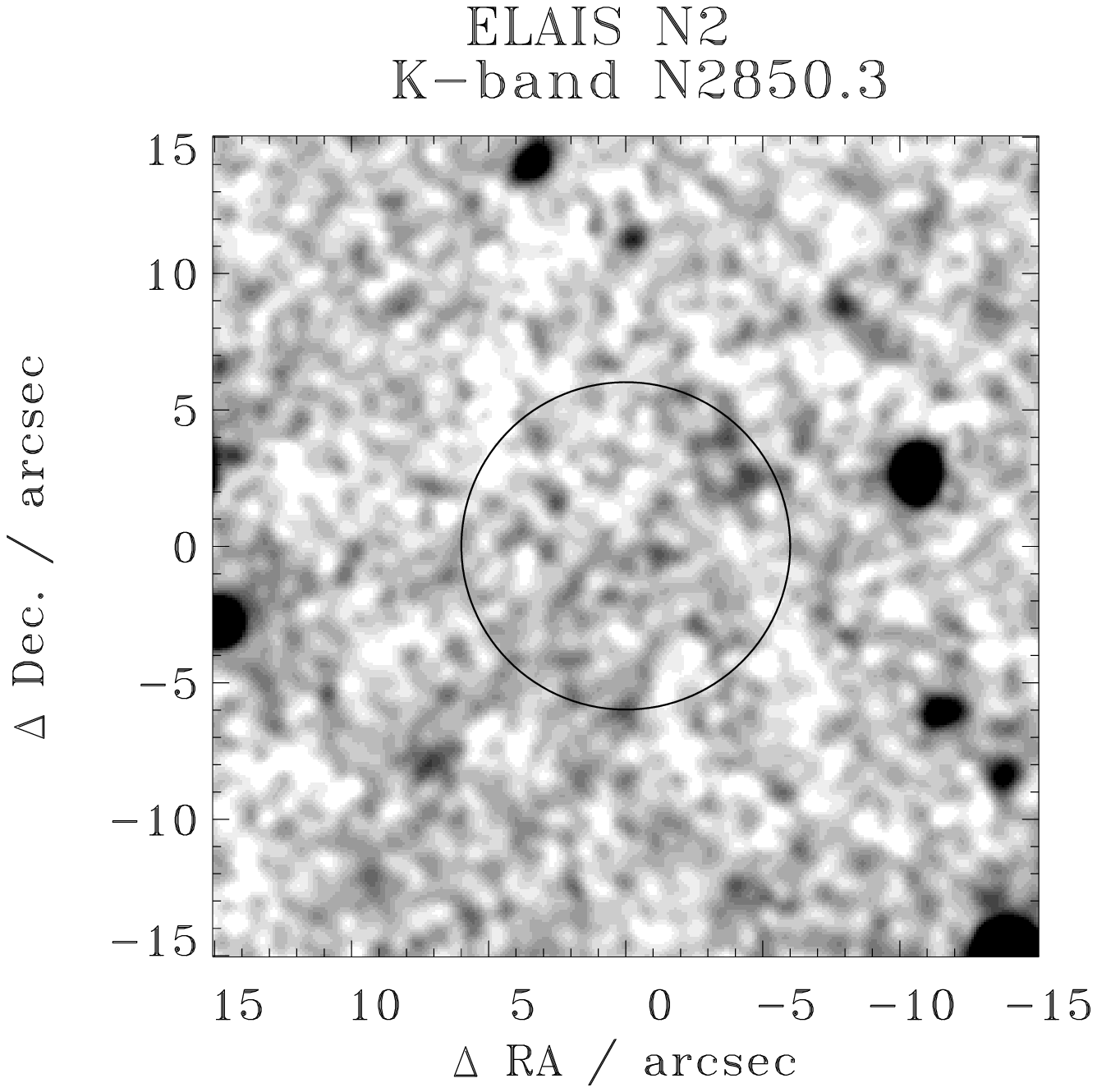}

\includegraphics{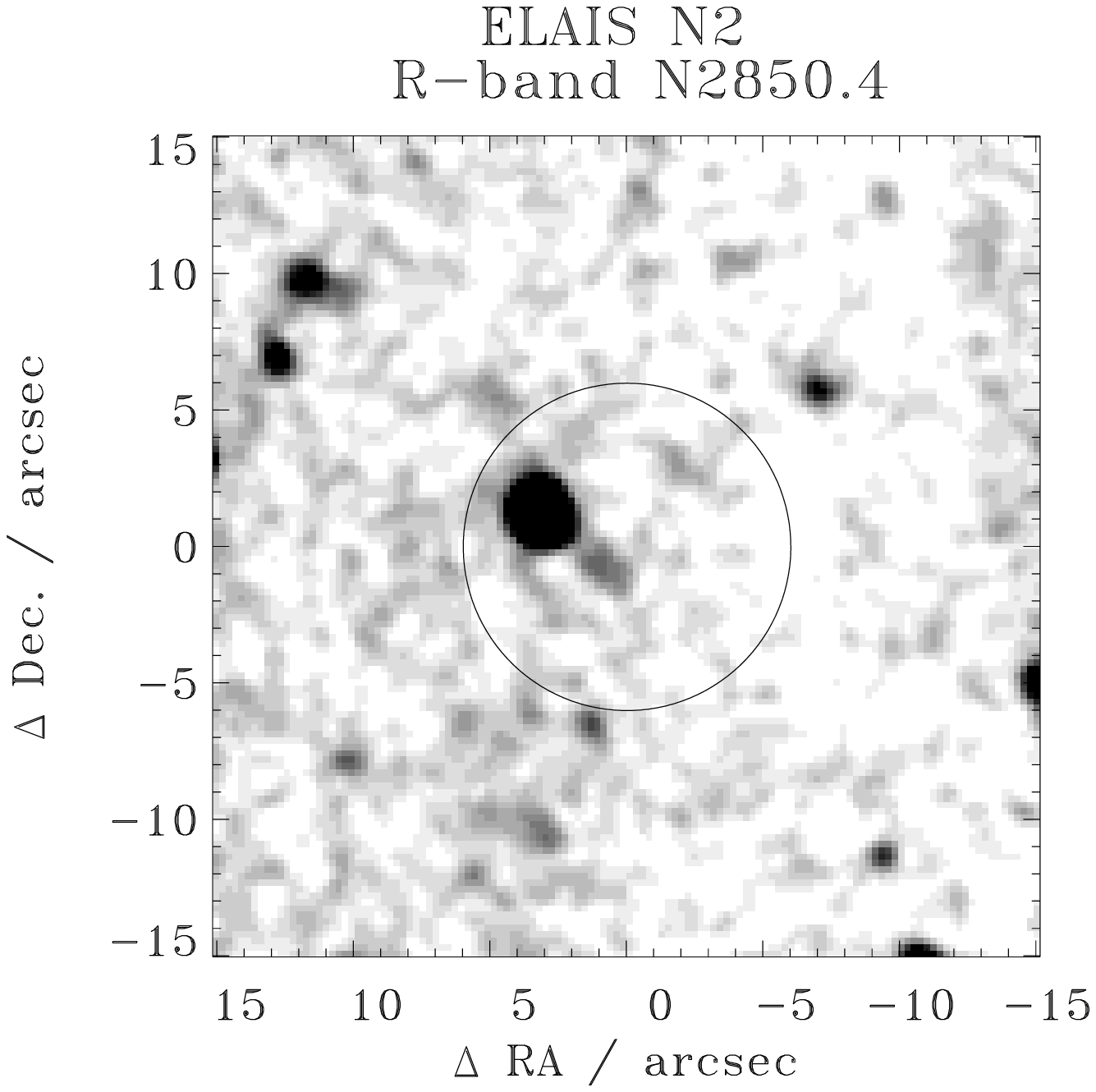}   
\includegraphics{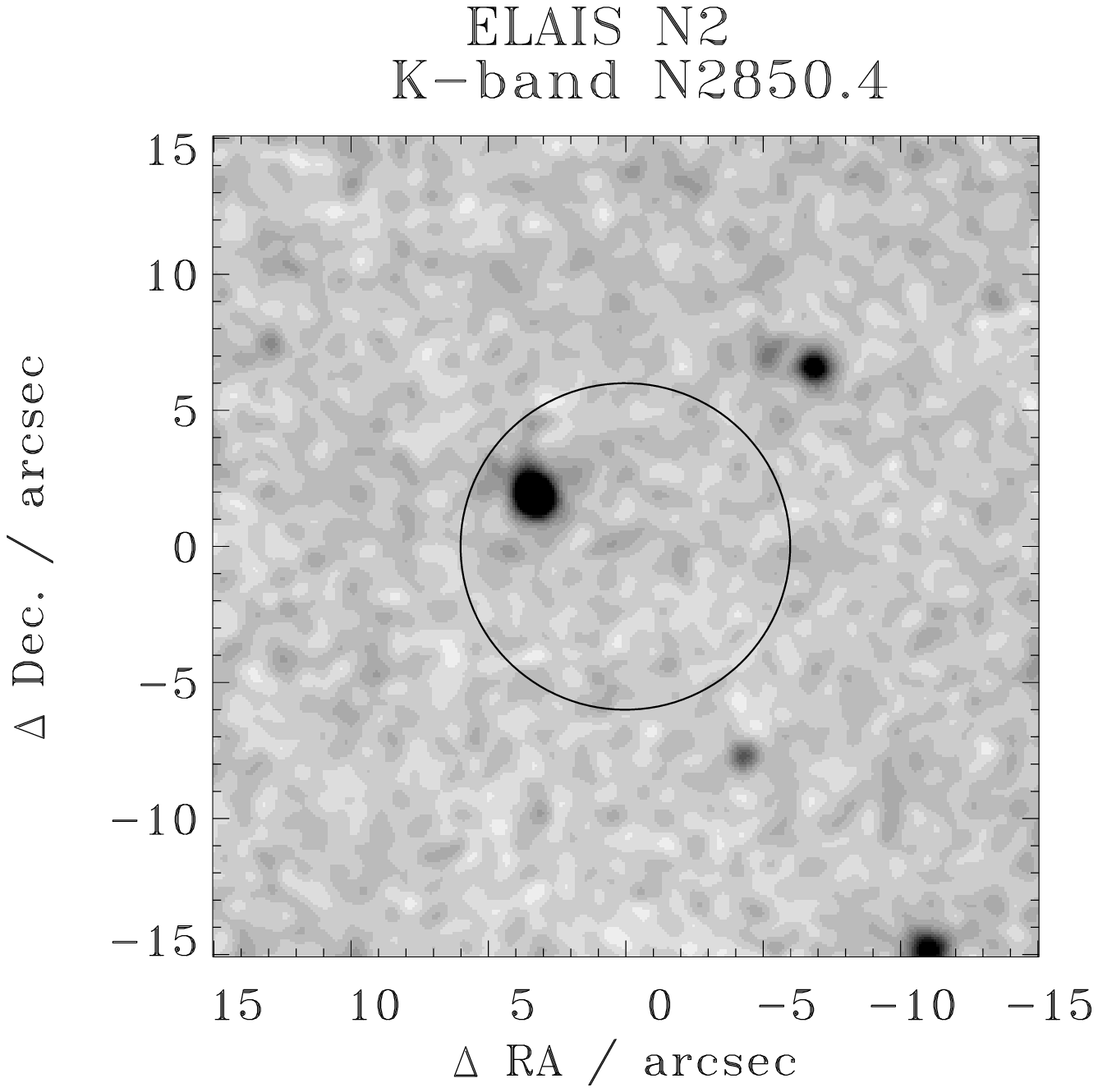}  
 
\includegraphics{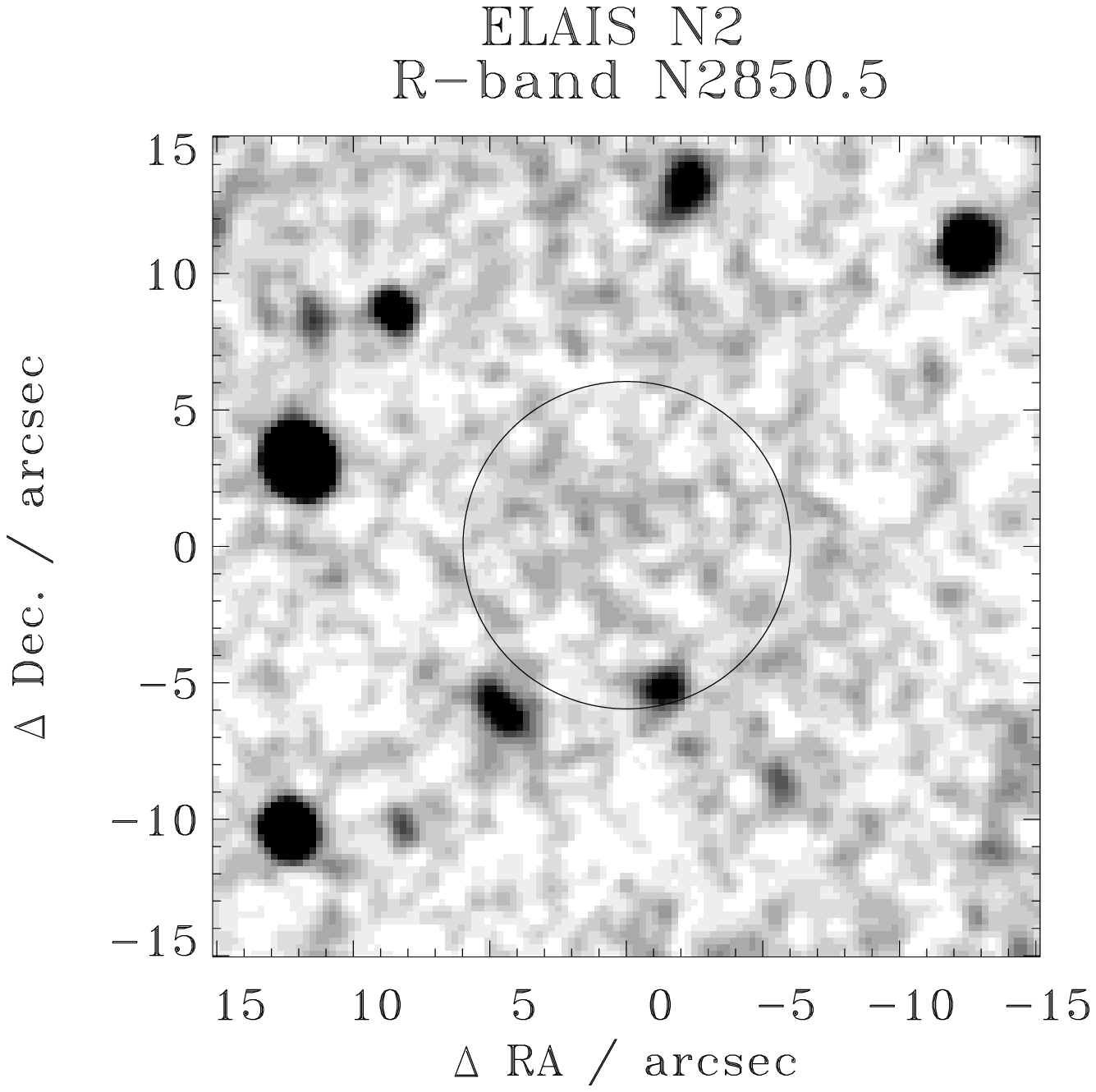} 
 
\includegraphics{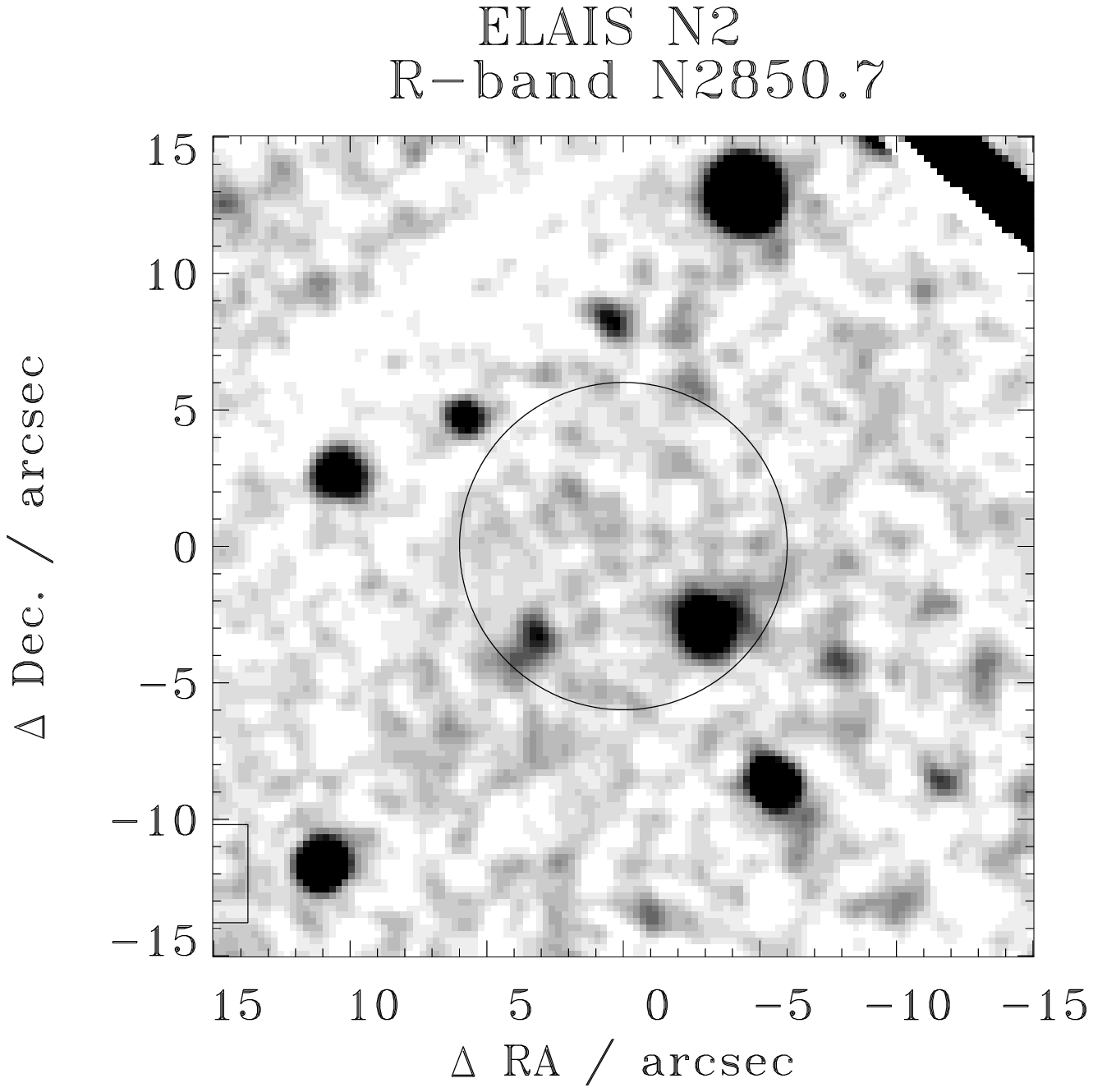}  
\includegraphics{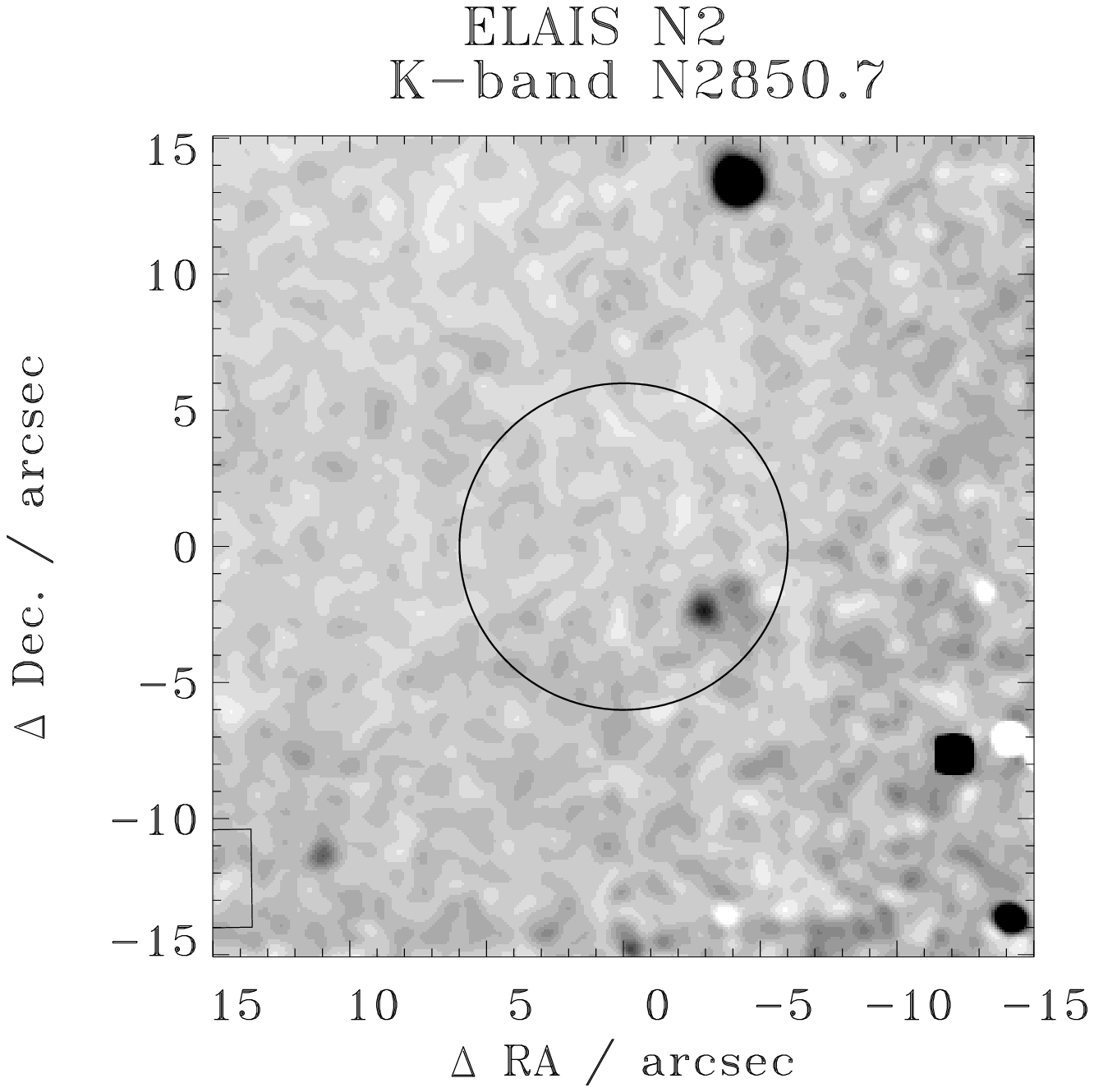}

 \caption[m]{$R$ and $K$-band $30 \times 30$ arcsec postage stamps, centred on  
the 850$\mu m$ positions of the ELAISN2 SCUBA sources,   
indicating potential optical and/or near-infrared  
counterparts to the sources uncovered at 850$\mu m$. 
The large circle in each figure has a radius of 6 arcsec, and defines  
the   
search radius adopted for the calculation of the statistical  
significance of each potential identification as described in section 4.1. 
Two sources. N2850.1 and N2850.2, have significant 450$\mu$m detections the positions of which, 
as in Figure 4, are indicated by circles with a radius of 3 arcsec. In the  
case of N2850.1, the position of the 450$\mu m$ source reinforces the  
likelihood that the statistically compelling optical/IR identification 
centred on the 850$\mu m$ position is correct. However, in the case of N2850.2 
the 450$\mu m$ position points towards one of the red objects seen only in the  
$K$-band image (near the western edge of the 850$\mu m$ error circle)  
as the most likely identification. 
\label{n2_postage}} 
 
 \end{figure*} 
 
\subsection{Near-infrared data}\label{kband} 
 
We have observed the central region of our Lockman Hole SCUBA survey  
area with the infrared camera INGRID mounted on the WHT,  
producing a single $4 \times 4$ arcmin $K$-band image.   
Figures 4 and 5 include $30 \times 30$ arcsec  
$K$-band postage stamps extracted from this image for  
the four 850$\mu$m detections which fall 
within this area.   
A smaller but substantially deeper $K$-band image taken with UFTI on UKIRT  
(Lutz et al. 2001) has revealed new faint possible near-infrared 
counterparts for LH850.1 and LH850.8.   
The $K$-band counterpart of LH850.1 has a magnitude of $K=21.4$ 
(within a 1.5-arcsec radius aperture), and is located less than 
1 arcsec from the refined position of the SCUBA source 
provided by its detection at 1.2mm by the IRAM PdB  
interferometer. There is no doubt that 
this faint, red and apparently complex object is the correct 
identification for the SCUBA source (Lutz et al. 2001).  For bright sources such as LH850.1 lensing would be a more common event if the counts are steep at brighter fluxes. 
 
The $K$-band image of LH850.8 is also of interest because  
it provides a third potential identification in addition to the two 
alternatives provided by the $I$-band image. 
This is a particularly complex source; the SCUBA error  
circle contains ROSAT X-ray source (LH850.8(1)) and VLA radio detection (LH850.8(2)), which 
appear to have distinct optical/IR counterparts, neither of which  
is necessarily a convincing identification for the SCUBA source. 
The radio source is VLA source 75 (according to the  
naming scheme of de Ruiter et al. 1997), and its VLA positional  
error box is shown in Figure \ref{lh_i_postage} to lie within the  
SCUBA error circle, coincident with one of the $K$-band sources. 
In fact de Ruiter et al. (1997) list this VLA source as a confident  
association with the ROSAT source 33.   
However, this association was based on the ROSAT PSPC centroid position  
which has poor spatial resolution.   
Subsequent observations with the HRI instrument, which has a much  
higher spatial resolution, indicate that ROSAT source 33 is associated with  
the upper $K$-band source as illustrated in Figure  
\ref{lh_i_postage} and not with the VLA detection (Lehmann et al. 2001).  Recent optical spectroscopy of the VLA source by Lehmann (priv. comm.) has revealed LH850.8(2) to be an emission line galaxy at $z=0.685$.  If the sub-mm source and the VLA source are indeed physically associated the spectrocopic redshift is consistent with broad range ($z=0.5 - 3$) derived from the Carilli and Yun (2000) indicator.

Our new, deep $K$-band image of this region has now  
revealed a third, faint ($K=20.2$) infrared source NE of the VLA source (denoted as LH850.3(3)), 
close to the 850$\mu m$ centroid but outside the VLA error box. 
The non-detection of this source in the $I$-band image  
indicates it is red, with $I-K > 4.28$ which, given our experience  
with LH850.1 strengthens our conviction  
that this is probably the true SCUBA identification, despite the  
fact that the $P_{E}$ statistic marginally favours the VLA source as the  
least likely chance association.  Table \ref{redshifts} lists the redshift limits of LH850.8 with no assumptions of the true counterpart and also the two spectroscopically derived redshifts.  Deep, high-resolution IRAM imaging will resolve this conundrum.  LH850.8 will be discussed further in Ivison et al. (2001, in preparation).             
 
A substantial fraction of our ELAIS N2 survey field has now also  
been mapped in the $K$-band, in this case with  
the UFTI camera on UKIRT. 
This dataset comprises 13 individual frames, each covering  
a field approximately $100 \times 100$ arcsec in size. Typical exposure  
times are 120 minutes per pointing, resulting in images  
which reach a 3-$\sigma$ detection limit of $K \simeq 21.5$ as measured 
with a 1.5-arcsec radius aperture.  
One frame containing two strong SCUBA detections has been imaged 
to an increased depth of $K \simeq 22$.   
Of the six bright SCUBA sources in the ELAIS N2 region,  
two  (N2850.1 and N2850.5)  
unfortunately lie outside the field covered by this deep $K$-band mosaic. 
$K$-band postage stamps covering $30 \times 30$ arcsec  
are provided for the remaining four sources in  
Figure 6. 
 
A number of SCUBA sources have now been convincingly shown to  
be associated with EROs with $R-K > 5$ (Smail et al. 1999, Frayer et al. 2000, Ivison et al. 2000).  
Consequently, particularly in the light of our own detailed study  
of LH850.1 (Lutz et al. 2001), we have good reason for taking 
particularly seriously any potential SCUBA identifications 
revealed in $K$-band images which transpire to be extremely  
faint, or undetected in the complementary optical data (Ivison et al 2001).  Three of four SCUBA galaxies from the lensing cluster survey were initially named as being associated with brighter optical counterparts until deeper near-IR imaging revealed the fainter, redder sources classing them as EROs (Smail et al 1999, Frayer et al. 2000). 
In addition to the cases of LH850.1 and LH850.8 discussed above, 
the possible $K$-band counterparts of three ELAIS N2 sources 
 are EROs with $R-K > 5.3$, $R-K > 5.8$ and $R-K > 6$  
for N2850.2(3), N2850.3(3) and N2850.7(3), respectively.   
Extrapolating the ERO number counts of Daddi et al. (2000) to the  
$K$-band limit of $21.5$ for $R-K > 6$ yields an estimated expected  
ERO density of 0.5 sq.arcmin$^{-1}$.  By chance we would thus  
expect only $\sim$0.1 EROs to fall within the 6 arcsec search radius in  
{\it one} of the eight possible fields for which we possess 
$K$-band imaging. Thus, the relative rarity of such 
red objects strengthens the argument that these EROs  
are indeed the correct identifications for the SCUBA sources. However, 
it must be acknowledged that the estimated ERO source density  
below K$\sim$20 is a source of large uncertainty.   
 
\subsection{X-ray data}\label{rosat} 
 
As mentioned above the SCUBA source  
LH850.8 is only 4.8 arcsec distant  from an X-ray source detected in the  
ROSAT deep survey (Hasinger et al. 1998, Lehmann et al. 2001) here named LH850.8(1). 
Deep $R$-band imaging and optical spectroscopy  
has been obtained for the optical counterpart of this  
X-ray source (No. 33 using Lehmann et al. (2000) naming scheme)  
by Lehmann et al. (2000). The spectrum displays narrow OII and  
NeV emission and a broad Mg II line, revealing  
this object to be an AGN at $z=0.9$.  
 
The proximity of this AGN to the SCUBA source LH850.8 is undeniably interesting 
but, as discussed above, neither this source nor the nearby VLA radio 
source (LH850.8(2)) can be unambiguously associated with the SCUBA source. 
At present we therefore have no compelling evidence for AGN activity  
in any of the 19 bright SCUBA sources considered here. 
 
However, deep Chandra observations in the field of Abell 370  
by Bautz et al. (2000) have revealed hard X-ray sources coincident with  
SCUBA sources and suggest that around 20$\%$ of the sub-mm population  
may have a significant contribution from an AGN component.   
 
Fabian et al (2000), Barger et al (2001a) and  Barger et al. (2001b) have found a similarly low sub-mm detection rate with X-ray samples; the contribution to the 850$\mu$m background light from hard X-ray sources estimated from these works to be lower; around 10$\%$.     
 
These results suggest that  
deeper X-ray imaging of the 8-mJy survey might also be expected 
to yield some convincing X-ray detections of our SCUBA sources. 
 
In fact a deep Chandra image of the ELAIS N2 field has now been obtained, 
and an analysis of the cross-correlation between the faint X-ray population  
and the SCUBA sources based on the 8-mJy survey will be the subject of  
a forthcoming paper by Almaini et al. (2001).

\begin{table*} 
\footnotesize 
\centering 
\caption[]{Positions of SCUBA sources and possible optical/infrared/radio/X-ray counterparts in the  
Lockman Hole E area.  
Column 2 shows detections close to the SCUBA centroid.  The positional errors are  
typically $\pm0.1-0.2^{\prime \prime}$ for the R/I/K-band sources, $\pm 2^{\prime \prime}$  
for the VLA sources and $\pm 2^{\prime \prime}$ for the SCUBA 450$\mu$m sources.   
The $P_{E}$-statistic (outlined in the text) quantifies the probability that  
the optical or infrared counterpart may be a chance coincidence  
(Section \ref{rband}). \label{lh_posn}}.     
\begin{tabular}{llccl}\hline 
Catalogue Name &   & $\alpha_{2000}$ &	$\delta_{2000}$          	&   Note   \\  [0.25ex]\hline 
      
LH850.1 &  SCUBA 850$\mu$m &    10 52  01.439 & +57 24 43.15   &   7.62 S/N   		\\	 
    &  SCUBA 450$\mu$m &    10 52  01.577 & +57 24 49.30    &  3.82 S/N   		\\ 
	   &   IRAM PdB 1.2mm   &    10 52  01.284 & +57 24 45.94   &           Lutz et al. 2001 \\ 
      &  K-band Peak     &    10 52  01.300 & +57 24 46.00    &	Lutz et al. 2001 \\ 
      &  VLA 1.4 GHz     &    10 52  01.249 & +57 24 45.88   &   Ivison et al. 2001 \\[0.75ex]  
LH850.2 &  SCUBA 850$\mu$m &   10 52 38.214 & +57 24 36.10   &   4.70 S/N  	              \\                       
      &  I-band Peak     &    10 52 38.280 & +57 24 40.93 &	$P_{E}$ = 0.40			\\[0.75ex] 
 
LH850.3 &  SCUBA 850$\mu$m &   10 51 58.272 & +57 18  01.14   &   4.93  S/N                 \\                      
	&I-band Peak     &   10 51 58.536  & +57 17  55.68  & $P_{E}$ = 0.46		\\  
 
LH850.4 &  SCUBA 850$\mu$m &   10 52  04.138 & +57 25 28.15   &   5.14  S/N                 \\                      
      &	 I-band Peak  (1)   & 10 52  04.440  & +57 25 29.96 & $P_{E}$ = 0.21			\\   
      &  K-band Peak  (1)   &  10 52  04.222 & +57 25 31.04 & $P_{E}$ = 0.60  				\\ 
      &	 I-band Peak  (2)   &  10 52  03.528  & +57 25 30.25 & $P_{E}$ = 0.22			\\ 
      &  K-band Peak  (2)   &  10 52  03.647 & +57 25 31.73   & 	$P_{E}$ = 0.54			\\ 
    &  K-band Peak  (3)   &  10 52  03.959 & +57 25 30.46   & 	$P_{E}$ = 0.55			\\ 
 
LH850.5 &  SCUBA 850$\mu$m &  10 51 59.341 & +57 17 17.65     &   4.50 S/N\\ 
   
LH850.6&  SCUBA 850$\mu$m &   10 52 30.582 & +57 22 11.59   &   4.24   S/N               \\                           
&  I-band Peak     &   10 52 30.792  & +57 22  09.59    & 	 $P_{E}$ = 0.23			\\

LH850.7 &  SCUBA 850$\mu$m &   10 51 51.456 & +57 26 35.12    &   4.34   S/N                 \\  
      &	 I-band Peak    &   10 51 51.984 &  +57 26 37.93  &  $P_{E}$ =  0.44 	\\

LH850.8&  SCUBA 850$\mu$m &    10 51 59.969 & +57 24 21.29   &   4.26 S/N             \\                     
      &  ROSAT HRI   (1)    &    10 52 00.0\phantom{00}   & +57 24 24.50   &	Lehmann et al. 2001    \\ 
      &  R-band Peak (1)	 &    10 52 00.0\phantom{00}   & +57 24 26.10   &	Lehmann et al. 2001   \\ 
      &  I-band Peak (1)    &    10 51 59.832 & +57 24 24.91   &   $P_{E}$ = 0.12	\\  
      &  K-band Peak (1)    &    10 51 59.905 & +57 24 25.30  &	   $P_{E}$ = 0.54			\\[0.5ex] 
 
      &	 VLA  1.4 GHz(2)    &    10 52 00.29\phantom{0}  & +57 24 20.3\phantom{0}    &	de Ruiter et al. 1997 \\  
      &	 I-band Peak (2)    &    10 52 00.192 & +57 24 19.69   &   $P_{E}$ = 0.11	\\  
      &  K-band Peak (2)  &    10 52  00.242 &   +57 24 19.97   &  $P_{E}$ = 0.28		   	\\[0.75ex]  
 
 	&  K-band Peak (3)  &    10 52  00.289 &   +57 24 22.97   & $P_{E}$ = 0.61		   	\\[0.75ex]  
 
LH850.11 &  SCUBA 850$\mu$m &   10 51 30.601 & +57 20 38.48   &   4.43   S/N                \\                       
      &  SCUBA 450$\mu$m &  10 51 30.949  &  +57 20 41.95    & 3.78  S/N            \\ 
	&I-band Peak     &   10 51 30.792  & +57 20 42.50  & $P_{E}$ = 0.40			\\

LH850.12&  SCUBA 850$\mu$m &   10 52  07.723 & +57 19  06.65   &   4.03   S/N                \\    
      &  I-band Peak     &   10 52  07.464 & +57 19  01.70    &  $P_{E}$ =  0.40 \\   
      &  VLA 1.4 GHz     &   10 52 07.49\phantom{0}   & +57 19  02.7    &   de Ruiter et al. 1997                 \\    
   &  I-band Peak     &   10 52  08.016 & +57 19  02.93    &  $P_{E}$ =  0.40\\

LH850.14 &  SCUBA 850$\mu$m &   10 52  04.298 & +57 26 59.16   &   4.64  S/N                  \\                        
   
LH850.16&  SCUBA 850$\mu$m &   10 52 27.080 & +57 25 16.23   &   4.15  S/N                 \\       
      &  I-band Peak     &   10 52 27.120  & +57 25 17.18& $P_{E}$ = 0.04			 \\	[0.75ex]   	     
 
LH850.18 &  SCUBA 850$\mu$m &   10 51 55.661 & +57 23 12.14   &   4.46   S/N                \\                         
&  I-band Peak (1) &    10 51 55.560 & +57 23 10.43   & $P_E$ = 0.15\\   
      &  I-band Peak (2) &   10 51 55.944 & +57 23 13.06  &   $P_E$ = 0.24  \\ [0.75ex]\hline

\end{tabular} 
\end{table*}

\begin{table*} 
\footnotesize 
\centering 
\caption[]{Positions of SCUBA sources and possible optical-infrared counterparts in the ELAIS N2 area.  
Column 2 shows detections close to the SCUBA centroid.  The positional errors are  
typically $\pm0.1-0.2^{\prime \prime}$ for the R/I/K-band sources, $\pm 2^{\prime \prime}$  
for the VLA sources and $\pm 2^{\prime \prime}$ for the SCUBA 450$\mu$m sources.   
The $P_{E}$-statistic (outlined in the text) quantifies the probability that  
the optical or infrared counterpart may be a chance coincidence  
(Section \ref{rband}). \label{n2_posn}}    
\begin{tabular}{llccl}\hline 
Catalogue Name & &   $\alpha_{2000}$ &	$\delta_{2000}$  	&   Note   \\  [0.25ex]\hline 
      
N2850.1 & SCUBA 850$\mu$m &    16 37  04.332  & +41 05 30.32   &   8.46  S/N    \\   
	& SCUBA 450$\mu$m &16 37  04.363 &   +41 05 28.64    &   4.24 S/N \\ 
      &  R-band Peak (1)     &   16 37  04.343  &  +41 05 31.24	&   $P_{E}= 0.06$         \\  
      &  I-band Peak (1)    &   16 37  04.331  &  +41 05 30.72	&   $P_{E}= 0.01$         \\  
	&  R-band Peak (2)    &  16 37 04.684 & +41 05 34.57    &    $P_{E}=0.93$  \\ 
	&  R-band Peak (3)    &  16 37 04.315  &+41 05 25.20  &   $P_{E}=0.94$  \\ 
 
N2850.2 &	SCUBA 850$\mu$m &    16 36 58.651  & +41 05 24.35   &   6.05 S/N \\  
	& SCUBA 450$\mu$m &	16 36 58.260  &  +41 05 23.70  &   3.60 S/N \\ 
	& R-band Peak    & Diff Spikes  		&&  \\  
	& I-band Peak (1)   & 16 36 58.662  & +41 05 25.71  &  $P_{E}=0.23$\\  
	& K-band Peak (1)   & 16 36 58.682  & +41 05 25.76  &  $P_{E}=0.33$\\  
	& I-band Peak (2)   & 16 36 58.731 & +41 05 21.87  &  $P_{E}=0.54$\\  
	& K-band Peak (2)   & 16 36 58.704 & +41 05 22.60  &  $P_{E}=0.45$\\  
	& K-band Peak (3)   & 16 36 58.198    &  +41 05 24.04   & $P_{E}=0.92$\\ 
	& I-band Peak (4)   & 16 36 58.884    &  +41 05 23.32   & $P_{E}=0.74$\\ 
	& K-band Peak (4)   & 16 36 58.872    &  +41 05 23.67   & $P_{E}=0.68$\\ 
	& I-band Peak (5)   & 16 36 58.963    &  +41 05 25.49   & $P_{E}=0.67$\\ 
 
N2850.3 &  SCUBA 850$\mu$m &    16 36 58.228  & +41 04 42.35   &   6.16 S/N  \\ 
      	&  R-band Peak (1)    &   16 36 58.083  &  +41 04 41.98  & $P_{E}=0.47$ \\ 
      	&  I-band Peak (1)    &   16 36 58.060  &  +41 04 41.55  & $P_{E}=0.50$ \\ 
	&  R-band Peak (2)  & 16 36 57.925  &  +41 04 42.39 &  $P_{E}=0.77$\\ 
	&  K-band Peak (3)  &	16 36 57.839 &  +41 04 44.79&  $P_{E}=0.94$ \\ 
 
N2850.4 &	SCUBA 850$\mu$m &    16 36 50.143  & +40 57 32.87   &   5.70 S/N \\  
	&  R-band Peak (1) & 16 36 50.180  &  +40 57 32.10	&  $P_{E}= 0.20$         \\  
	&  R-band Peak (2) & 16 36 50.435  &  +40 57 34.46 	&  $P_{E}= 0.34$         \\  
	&  I-band Peak (2) & 16 36 50.433  &  +40 57 34.54 	&  $P_{E}= 0.32$         \\  
	&  K-band Peak (2) & 16 36 50.424  &  +40 57 34.88      &  $P_{E}= 0.44$  \\ 
	&  R-band Peak (3) & 16 36 50.258  &  +40 57 32.64      &  $P_{E}= 0.31$  \\ 
 
N2850.5 &	SCUBA 850$\mu$m &    16 36 35.624  & +40 55 57.86   &   5.64 S/N \\  
	&  R-band Peak  & 16 36 35.518  &  +40 55 53.05		& $P_{E}= 0.85$          \\  
	&  I-band Peak  & 16 36 35.528  &  +40 55 52.87		& $P_{E}= 0.88$          \\

N2850.7 &	SCUBA 850$\mu$m &    16 36 39.415  & +40 56 38.37   &   5.37 S/N \\  
	&  R-band Peak (1) &  16 36 39.155 &  +40 56 35.93   &  $P_{E}= 0.56$  \\ 
	&  I-band Peak (1) &  16 36 39.174 &  +40 56 35.87   &  $P_{E}= 0.45$  \\ 
	& K-band Peak (1)  &  16 36 39.144 & +40 56 35.96   &  $P_{E}= 0.70$  \\ 
	&  R-band Peak (2) &  16 36 39.713 & +40 56 35.50&     $P_{E}= 0.84$\\ 
	&  I-band Peak  (3)   &  16 36 39.052 & +40 56 36.96  &   $P_{E}= 0.76$ \\ 
	& K-band Peak  (3)   &  16 36 39.049 & +40 56 36.53  &   $P_{E}= 0.92$ \\ [0.5ex]\hline 
 
\end{tabular} 
\end{table*}

\section{Discussion}\label{disc} 
 
Determining the redshift distribution of the sub-mm selected galaxy  
population is now regarded as a key goal in observational cosmology. 
This is of importance for assessing the contribution of dust enshrouded 
star-formation activity to overall star-formation density at  
high redshift, and for determining whether the massive starbursts which power 
these objects are spread throughout much of cosmic history, 
or mainly confined to a relatively short-lived epoch.  
Division of sub-mm selected samples into (even crude)  
redshift bands will also be of importance for refining current measurements 
of sub-mm source clustering (Scott et al. 2001), measurements which have 
the potential to settle the issue of whether bright sub-mm sources are 
the high-redshift progenitors of present-day massive ellipticals. 
 
Given the growing evidence that many, and perhaps most sub-mm sources 
have very faint, often red optical/IR identifications, it seems clear 
that the measurement of  spectroscopic redshifts for significant numbers  
of SCUBA sources will be a long term project. Indeed for some sources 
such measurements may not be feasible until the advent of deep infrared 
spectroscopy with NGST, or broad-band millimetre spectroscopy with ALMA 
or the LMT. 
 
Therefore, as stressed in the introductory section of this paper, while not 
losing sight of the ultimate goal of spectroscopic redshifts, it is important 
to recognize what can be learned about the redshifts of sub-mm sources 
using currently operational facilities. In particular it seems likely  
that, to first order, the basic redshift distribution of the sub-mm source 
population can be derived from broad-band radio-submm  
photometric constraints, coupled with the study of potential  
counterparts revealed by deep optical and near-infrared imaging. 
However, the usefulness of such techniques has until now been hampered 
by the lack of a substantial and unbiased sample of submm-selected  
sources of sufficient luminosity to allow detection of the  
majority of the 850$\mu m$ sources at other  
wavelengths ({\it e.g.} radio, mm and far-infrared wavelengths). 
 
It is of course important to recognize that a meaningful sample of 
apparently bright sub-mm sources has been provided by SCUBA observations of 
lensing clusters (Smail et al. 1997), and that the extensive follow-up of 
these sources (Ivison et al.\ 1998, 2000; Frayer et al.\ 1998, 1999, 2000) 
has produced great strides in our knowledge of the sub-mm population.  The 
lensing strategy is not without problems, however --- for example the small 
fields of view severely hamper measures of the clustering properties (if 
any) of SCUBA sources.

With completion of 850$\mu m$ source extraction  
from the 8-mJy SCUBA survey (Scott et al. 2001) we now, for the first time, 
possess the required statistically meaningful and  
unbiased sample of bright sub-mm sources. The results reported here 
thus represent an important step towards measuring the redshift distribution 
of the luminous sub-mm galaxy population. 
 
The key result of this work is that the SED-based redshift constraints, in  
particular the more powerful constraints provided by the combination  
of the 850$\mu m$, 450$\mu m$ and 20-cm data, all point to the same  
conclusion that essentially all of the 8-mJy sources lie at $z > 1$ and  
that at least half appear to lie at $z > 2$. At the same time 
the upper limits on redshift, where available, do not violate  
these minimum redshift constraints but suggest that not many of the 
sources are likely to lie at very extreme redshifts ($z > 4$). 
We have also demonstrated that candidate optical and/or near-infrared  
counterparts, while rarely offering unambiguous identifications given the 
current positional uncertainties, are certainly consistent with the  
SED-based redshift estimates. 
 
These redshift constraints may appear crude, but nonetheless  
are potentially very significant. In particular they confirm that 
most of the star formation which occurs in very extreme  
starbursts (SFR $> 1000$~${\rm M_{\odot} yr^{-1}}$) is confined  
to the first 2-3 Gyr of the history of the universe. The stellar populations 
produced by this population must therefore appear highly coeval 
and typically $> 10$ Gyr old by the  
present day, strengthening the argument that high-redshift sub-mm sources 
are the progenitors of present-day evolved ellipticals. Moreover, these 
 redshift bounds also confirm the plausibility of the redshift  
ranges adopted by Scott et al. (2001) for the estimation 
comoving number density of luminous dust-enshrouded starbursts in the young 
universe. This calculation which yields a value comparable to the present-day comoving number density 
of luminous ($L > 3L^{\star}$) ellipticals,  $\simeq 1 \times 10^{-5}  
{\rm Mpc^{-3}}$, provides further  
(albeit circumstantial)  
support for the plausibility of the evolution of faint sub-mm source into present day massive elliptical.

The other most interesting result of the analysis presented in this 
paper is the tantalizing suggestion that a large fraction of  
very luminous sub-mm sources may transpire to be associated with faint EROs. 
Specifically, while at present we only possess deep $K$-band images  
of 8 of the 19 most significant 850$\mu m$ sources, 5 out of these 8 images 
have revealed a potential ERO counterpart to the SCUBA source.  
In the case of LH850.1 the validity of this association has now been 
demonstrated beyond doubt, and the relative rarity of EROs in the field   
adds further credence to the other possible ERO associations. 
Thus, while the overall SCUBA population may contain  
a wide range of different classes of object, it seems  
possible that a substantial fraction of the brightest SCUBA sources 
sampled by the 8-mJy survey may well be EROs. 
Deeper VLA observations of our survey fields have the potential 
to yield much more accurate positions for a substantial fraction  
of the 8-mJy sources, and are thus expected to clarify  
which of the potential optical/IR identifications highlighted  
in the previous section can indeed be reliably  
associated with the SCUBA sources (Ivison et al. 2001). 
 
\section{Conclusions} 
 
In summary the main results and conclusions of the SCUBA 8-mJy survey are:\\

\noindent$\bullet$ All of the faint SCUBA sources detected in this survey lie at $z > 1$ and at least 50$\%$ appear to lie at $z > 2$.\\ 
\noindent$\bullet$ The SED-derived redshift limits and ranges agree with the extreme star formation rates ($>1000M_{\odot}yr^{-1}$) for SCUBA sources calculated in Scott et al. (2001)  \\ 
\noindent$\bullet$ For the SCUBA sources for which we have deep near infrared data there are strong indications that EROs and faint SCUBA sources are physically associated.\\

\section*{Acknowledgements} 
The authors wish to thank Thomas Greve for the reduction of the $I$-band  
imaging of the Lockman Hole field, along with Dieter Lutz and Andrew Baker for  
undertaking and reducing the 1.2mm photometry measurements  
made with the IRAM 30m telescope.  The authors would also like to thank Ingo Lehmann for additional spectra from the follow-up of ROSAT deep survey.  Matthew Fox, Andreas Efstathiou, Susie Scott, Stephen Serjeant, Bob Mann, Seb Oliver and Chris Willott all acknowledge 
the support of PPARC during the course of this project. Omar Almaini 
acknowledges the support of the Royal Society, while James Dunlop acknowledges 
the enhanced research time afforded by the award of a PPARC Senior Fellowship. 
The JCMT is operated by the Joint Astronomy Centre on behalf of the  
UK Particle Physics and Astronomy Research Council, the Canadian National  
Research Council and the Netherlands Organization 
for Scientific Research. The UKIRT is operated by the Joint Astronomy Centre  
on behalf of the UK Particle Physics and Astronomy Research Council. 
The WHT is operated by the Isaac Newton Group on behalf of the  
UK Particle Physics and Astronomy Research Council and the Netherlands  
Organization for Scientific Research.

\end{document}